\documentclass[a4paper,10pt,twocolumn]{article}
\usepackage{jpc}
\usepackage{graphicx}
\usepackage{color}

\def\mys#1{{\mbox{\scriptsize{#1}}}}    

\def\elj{\epsilon}             
\def\feff{f_{\mys{eff}}}        
\def\gp{g_{\mys{P}}}            
\def\KB{k_{\mys{B}}}            
\def\KBT{k_{\mys{B}} T}         
\def\lb{\ell_{\mys{B}}}         
\def\np{n_{\mys{P}}}            
\def\NM{N_{\mys{m}}}            
\def\NP{N_{\mys{P}}}            
\def\NC{N_{\mys{c}}}            
\def\QP{Q_{\mys{P}}}            
\def\rpp{r_{\mys{PP}}}          
\def\rp{r_{\mys{P}}}            
\def\rhop{\rho_{\mys{P}}}       
\def\rhom{\rho_{\mys{m}}}       
\def\rhoc{\rho_{\mys{c}}}       
\def\RE{R_{\mys{E}}}            
\def\RG{R_{\mys{G}}}            
\def\RH{R_{\mys{H}}}            
\def\ls{l_{\mys{S}}}            
\def\V#1{{\bf #1}}              

\renewcommand{\textfraction}{0.0}
\renewcommand{\topfraction}{1.0}
\renewcommand{\bottomfraction}{1.0}

\begin{document}
\unitlength 1cm  
\renewcommand{\textfraction}{0.0}
\renewcommand{\topfraction}{1.0}
\renewcommand{\bottomfraction}{1.0}


\title{Single chain properties of polyelectrolytes in poor solvent}

\author{Hans J{\"o}rg Limbach, Christian Holm\\
{\em Max-Planck-Institut f\"{u}r Polymerforschung,
Ackermannweg 10, 55128 Mainz,
Germany}
}



\maketitle

\begin{abstract}
  
  Using molecular dynamics simulations we study the behavior of a
  dilute solution of strongly charged polyelectrolytes in poor
  solvents, where we take counterions explicitly into account. We
  focus on the chain conformational properties under conditions where
  chain-chain interactions can be neglected, but the counterion
  concentration remains finite.  We investigate the conformations with
  regard to the parameters chain length, Coulomb interaction strength,
  and solvent quality, and explore in which regime the competition
  between short range hydrophobic interactions and long range Coulomb
  interactions leads to pearl-necklace like structures.  We observe
  that large number and size fluctuations in the pearls and strings
  lead to only small direct signatures in experimental observables
  like the single chain form factor. Furthermore we do not observe the
  predicted first order collapse of the necklace into a globular
  structure when counterion condensation sets in.  We will also show
  that the pearl-necklace regime is rather small for strongly charged
  polyelectrolytes at finite densities. Even small changes in the
  charge fraction of the chain can have a large impact on the
  conformation due to the delicate interplay between counterion
  distribution and chain conformation.

\end{abstract}

\section{Introduction}

Polyelectrolytes (PEs) are polymers which have the ability to
dissociate charges in polar solvents resulting in charged polymer
chains (macroions) and small mobile counterions~\cite{barrat96a}.
Because of their great relevance in technical applications as well as
in molecular biology they enjoy an increasing attention in the
scientific community~\cite{hara93a,schmitz93a,foerster95a,holm01a}.
The combination of macromolecular properties and long-range
electrostatic interactions results in an impressive variety of
phenomena which makes these systems interesting from a fundamental
point of view.

In this paper we focus on the special case of polyelectrolytes under
poor solvent conditions.  The reason for this is that a large number
of polymers are based on a hydrocarbon backbone for which water is a
very poor solvent. The solubility in water is often only given due to
their charged side groups. Important examples are sulfonated
poly-styrene (PSS), poly-methacrylic acid (PMA), DNA and virtually all
proteins.  The poor solvent conditions give rise to a competition
between the attractive interactions due to the poor solubility of the
backbone and the electrostatic repulsion of the PE charges.  This can
lead to elongated strings of locally collapsed structures (pearls),
commonly called pearl-necklaces.  Such necklace conformations have
been proposed on the basis of fluorescence studies~\cite{ghiggino85a}.
They have also been predicted in terms of scaling arguments in
refs.~\cite{kantor94a,kantor95a,dobrynin96a} for a weakly charged
single chain PE. Later this has been extended to strongly charged
chains and finite
concentrations~\cite{schiessel98a,schiessel99a,dobrynin99a,dobrynin01a}.
Also other theoretical approaches support the existence of necklace
conformations~\cite{solis98a,picket01a,migliorini01a}.  The scaling
approach of ref.~\cite{dobrynin96a} was supplemented with a Monte
Carlo simulation of a single chain that shows a cascade from one to
two to three globules with increasing strength of the electrostatic
repulsion. In their study the polyelectrolyte was weakly charged and
every monomer carried a fractional charge. The chain was studied in
the infinite dilution limit, where counterions have not to be taken
into account.  The formation of the necklace structure is due to the
Rayleigh instability of a charged droplet which leads to a split once
a critical charge is reached. The size of the pearls is determined by
the balance between electrostatic repulsion and surface tension. The
distance between two pearls is governed by the balance of the
electrostatic pearl-pearl repulsion and the surface tension.

Some aspects of the theoretical pearl-necklace picture have been
confirmed by simulations using the Debye-H\"uckel
approximation~\cite{lyulin99a,chodanowski99a} and with explicit
counterions~\cite{micka99a,micka00a,limbach01a,limbach02a,holm03a,limbach02c}.
However, there is up to now no clear experimental proof for the
existence of necklace chains. Conformational chain properties have
been observed which seem to be consistent with the necklace
picture~\cite{ghiggino85a,spiteri97a,carbajal00a,heinrich01a,aseyev01a,minko02a,lee02a}.

Also the scaling of the peak position of the structure factor $q^*$
with the polymer density $\rho$ has been thoroughly investigated which
however reveals only properties that occur in the semi-dilute regime
of interacting chains.  For good solvent chains in the semi dilute
regime the exponent $\beta$ in the scaling relation $q^* \propto
\rho^\beta$ was measured to be $\beta=0.5$ which is also the
theoretically predicted value~\cite{degennes76a} whereas for poor
solvent polyelectrolytes no single value for the exponent seems to
exist.  The experimental values vary between $\beta=0.3$ and
$\beta=0.5$~\cite{essafi95a,heitz99a,waigh01a,baigl_privcomm} and show
a dependence on the charge fraction of monomers which is implicitly
also responsible for the poor solvent parameter of the chain. The
theoretical predictions show (for fixed poor solvent and fixed charge
fraction) also a complicated transition from a $\beta=\frac{1}{2}$
regime via a $\beta=\frac{1}{3}$ into a crossover
scaling~\cite{dobrynin99a,dobrynin01a}, whereas recent simulations
measured a constant exponent $\beta=\frac{1}{3}$ for the whole
concentration range from semi-dilute up into the dense
regime~\cite{limbach02c}, so its fair to say that things are far from
being well understood.

However, even for the dilute concentration regime, the situation can
be more complicated as envisioned in the scaling approach, since the
entropy of the chain and of the counterions as well as the
electrostatic interaction between counterions and the PE charges have
to be taken into account.  In our previous shorter communications we
reported large conformational
fluctuations~\cite{limbach02a,holm03a,limbach02c} in the
pearl-necklace structures and showed that they were responsible for
the absence of strong signatures in the force extension relation and
in the form factor.

The aim of the present simulational study is a more detailed
investigation of the structure of strongly charged polyelectrolytes in
poor solvent in the dilute concentration regime where the chain-chain
interaction is weak so that one deals effectively with single chain
properties. Our focus will be a thorough data analysis of the observed
pearl-necklace conformations. To this end we had to develop a new
cluster recognition algorithm that is capable to characterize these
interesting conformations automatically from our simulated
configurational data. Moreover we look at the stability of the
pearl-necklace conformations in the presence of condensed counterions,
and perform a study of the Coulomb induced collapse transition.  We
will compare our results with predictions from scaling theory and will
discuss the validity range of the scaling approach for strongly
charged chains at finite density. We then attempt to give a
preliminary phase diagram for the systems studied. The last part will
be devoted to some experimentally accessible observables, like
characteristic chain size ratios and form factor. This should be
helpful in supporting the evaluation of experimental data in terms of
pearl-necklace signatures.

Our paper is organized as follows: After explaining the used
simulation method in sec.~\ref{sec:simu} and giving a short overview
over the simulated systems in sec.~\ref{sec:simusystems} we will
discuss our data analysis methods in sec.~\ref{sec:obs}. In
sec.~\ref{sec:scaling} we will compare our results in the
pearl-necklace regime with predictions made by scaling theories. In
the next sec.~\ref{sec_fluctuations} we will quantify the role of
fluctuations, and describe in sec.~\ref{sec_pase_space} our view of
the Coulomb induced collapse transition and a preliminary phase
diagram for the range of our simulation parameters. We then discuss in
detail some measurable observables in sec.~\ref{sec_formfac} and
finally end with our conclusion in sec.~\ref{sec:conclusion}.  In the
appendix we give a detailed overview of the simulated systems,
including a list of parameters and results for selected basic
observables.

\section{Simulation method}\label{sec:simu}

Our model of a PE solution and our molecular dynamics approach has
been described previously in detail
in~\cite{micka99a,limbach01c,limbach02a}.  It consists of $\NP$
flexible bead-spring chains with $\NM$ monomers and $\NC$ counterions
which are located in a cubic simulation box of length $L$ with
periodic boundary conditions. A fraction $f$ of the monomers is
monovalently charged ($\nu_{\mys{m}}=1$). Thus the total charge per
chain is $\QP = f \NM$. The number of counterions which are also
monovalently charged ($\nu_{\mys{c}}=-1$) is chosen such that the
overall system is electrically neutral. Densities are given either as
monomer density $\rhom=\NP \NM / L^3$ or charge density $\rhoc= 2 \, f
\NP \NM / L^3$.

All particles interact via a Lennard-Jones (LJ) potential
$4\,\epsilon[(\frac{\sigma}{r})^{12}-(\frac{\sigma}{r})^{6}-c]$ for
distances $r < R_c$ and zero elsewhere.  The constant $c$ is chosen
such that the potential value is zero at the cutoff $R_c$, and
$\epsilon$ is a measure of the solvent quality.  Monomers interact up
to $R_c=2.5\sigma$ giving them a short range attraction which can be
tuned by changing the value of $\epsilon$.  The counterions interact
via a purely repulsive LJ interaction with $R_c=2^{1/6}\sigma$.  The
units of length, energy and time are $\sigma$, $\epsilon$, and $\tau$,
respectively.

The chain monomers are in addition connected along the chain by the
finite extendible nonlinear elastic (FENE) bond potential of the form
$- 14 \KB T \ln [1- (r /(2\sigma))^2 ]$ which results in an average
bond length $b \approx 1.1 \sigma$.

Charged particles with charges $q_i$ and $q_j$ at separation $r_{ij}$
interact via the Coulomb energy $\KB T \lb q_i q_j / r_{ij}$ where the
Bjerrum length is defined as
$\lb=e^{2}/(4\pi\epsilon_{S}\epsilon_{0}\KB T)$ ($e$: unit charge,
$\epsilon_{0}$ and $\epsilon_S $: permittivity of the vacuum and of
the solvent).  The Coulomb interaction was calculated with the
P3M-algorithm~\cite{deserno98a,deserno98b}, tuned to force accuracies
which are much higher than the thermal noise level.

A velocity Verlet algorithm with a standard Langevin thermostat is
used to integrate the equation of motion~\cite{grest86a} (friction
coefficient $\Gamma = \tau^{-1}$, time step $\Delta t=0.0125\tau$).
Thus the solvent is only implicitly present via its permittivity
$\epsilon_S$, the friction constant $\Gamma$ and the solvent quality
parameter $\epsilon$ in the LJ potential.

The simulation time after equilibration for all systems was at least
$100$ times the measured correlation time for the end-to-end distance
$\RE$ and the chains centers of mass diffused at least several radii
of gyration $\RG$. The osmotic pressure $p$ was measured to be always
positive and additional simulations over a large density
range~\cite{limbach02c} showed that the $p V$ diagram is convex at all
densities, thus our simulations are stable, reach true thermal
equilibrium, and reside in a one phase region.  The volume density
inside the pearls does not exceed $0.47$ which is below the glass
transition. We therefore are certain that also the pearl formation and
restructuring was observed in equilibrium.

\section{Simulated systems}\label{sec:simusystems}

We have simulated our poor solvent polyelectrolytes mainly as a
function of the parameters chain length $\NM$, solvent quality $\elj$,
strength of the electrostatic interaction $\lb$ and the charge
fraction $f$. All simulations are performed in dilute solution such
that the interaction between the chains is small. For the system with
the longest chains ($\NM=478$) we have a chain extension $\RE \approx
60\sigma$ and a chain-chain separation $r_{cc}\approx 252\sigma$ which
was calculated for a random packing of spheres according to

\begin{equation}
r_{cc} \approx 1.28 \left( \frac{3 \pi}{4} \frac{L^3}{\NP} \right)^{1/3} \, .
\end{equation}

The screened renormalized monopole interaction
$U_{\mys{cc}}^{\mys{DH}}$ between two chains can be estimated for this
case to be of the order $\KB T$.
%
For our estimate we have used a crude approximation on the
Debye-H\"uckel level. The effective chain charge $Q_{\mys{P,eff}}
\approx 64$ is calculated by using the counterions within a shell of
$3\sigma$ around the polyelectrolyte to renormalize the bare charge
$\QP=160$.  The density of the free counterions $\tilde{\rhoc}$ leads
a screening constant $\kappa = \sqrt{4 \pi \lb \tilde{\rhoc}} \approx
0.009\sigma^{-1}$ and as the interaction potential we take
$U_{\mys{cc}}^{\mys{DH}} = \lb Q_{\mys{P,eff}}^2 e^{-\kappa r_{cc}} \,
r_{cc}^{-1} \approx 2.3 \KB T$.  For this estimate one has to keep in
mind that both the used practical definition of the effective charge
as well as the screening concept can not be founded on physical
principles.

Details about the used parameters and measurements of some basic
observables for all simulated systems can be found in appendix A. We
have grouped the simulations into series depending on the investigated
parameters.

The chain length dependence is studied in simulation series which
differ in the line charge density $f$. For series A1 we use a charge
fraction $f=\frac{1}{3}$, and the chain length is varied over one
decade in steps of $48$ from $\NM=48$ to $478$. Series A2 and A3 are
performed at $\NM=100$, $200$ and $300$ with $f=\frac{1}{2}$ at
slightly different densities.

To study the dependence on the solvent quality we have varied the
short range attraction by changing the LJ parameter $\elj$ between
$0.0 \KB T$ and $2.0 \KB T$ for medium sized chains of length
$\NM=238$ (series B). The $\Theta$-point for this model was determined
to be at $\elj(\Theta)=0.34\KB T$~\cite{micka99a}.  For most of the
other simulations (series A1, A2, A3, C1, C2 and C3) we used $\elj =
1.75 \KB T$ which is thus deep in the poor solvent regime.  This value
is chosen for practical reason. In this regime we have found
relatively large and stable pearl-necklace conformations which are
more easy to investigate.

The effect of the Coulomb interaction is studied via changing the
Bjerrum length $\lb$ and the charge fraction $f$. We simulated chains
with length $\NM=199$ in three series C1, C2 and C3 with charge
fractions $f=1$, $f=\frac{1}{2}$ and $f=\frac{1}{3}$ respectively.
They are performed at the same charge density
$\rhoc=5\times10^{-5}\sigma^{-3}$ but they contain different numbers
of counterions $\NC = f \NM \NP$ corresponding to the number of
charges on the PEs. All three series start out at $\lb=0\sigma$.  In
Series C1 $\lb$ ranges up to $10\sigma$, in series C2 and C3 up to
$9\sigma$. In addition there is a simulation set with shorter chains
$\NM=94$ with $f=\frac{1}{3}$ and a different solvent parameter
$\elj=1.5\KB T$ which we only use for the phase diagram (series C4).

The last series contains four simulations with the same value for the
scaling variable $\lb b^{-1} f^2=0.25$ but different $f = 1$,
$\frac{1}{2}$, $\frac{1}{3}$ and $\frac{1}{4}$ (series D).

\section{Observables and data analysis}\label{sec:obs}

In this section we define our measured observables and explain in
detail how we analyzed our simulated PE conformations. Especially we
present the cluster recognition algorithm which we used to
automatically classify different pearl-necklace structures.

We denote the position of monomer $i$ with $\V{r}_i$ and the distance
between two particles $i$ and $j$ with $r_{ij}$. The center of mass
for the chain is then $\V{R}_s=\frac{1}{\NM} \sum_{i=1}^{\NM} \V{r}_i$
and the center of mass coordinates are $\V{x}_i=\V{r}_i-\V{R}_s$.  For
the chain extension we use the end-to-end distance

\begin{equation}
\RE^2 = \left( \V{r}_1 - \V{r}_{\NM} \right)^2 \mbox{,}
\end{equation}

the radius of gyration

\begin{equation}
\RG^2 = \frac{1}{\NM} \sum_{i=1}^{\NM} |\V{x}_i|^2
\end{equation}

and the inverse hydrodynamic radius

\begin{equation}
\RH^{-1} = \frac{1}{\NM(\NM-1)} \sum_{i \neq j}^{\NM} \frac{1}{r_{ij}} \mbox{ .}
\end{equation}

Note that this definition corresponds to the short time diffusion
behavior of polymers. For more information on this topic see
ref.~\cite{liu03a}.

For a first structure classification we use two characteristic ratios
between the different chain extension observables. The first
characteristic ratio is defined as $r=\left( \RE / \RG \right)^2$.
The second characteristic ratio is $\alpha= \RG / \RH$ which has the
advantage of being experimentally accessible~\cite{schweins01a}.

We will also compute the spherically averaged form factor, sometimes
also called single chain structure factor, $S_1(q)$ which can be
measured in scattering experiments:

\begin{equation}
   S_1(q) = \frac{1}{\NM} \sum_{i=1}^{\NM} \sum_{j=1}^{\NM}
\frac{\sin(q r_{ij})}{q r_{ij}} \, \mbox{.} \label{formfac_sph} 
\end{equation}

To describe the counterion distribution around polyelectrolytes we use
an integrated counterion distribution $P(r)$.  The distance $d_i$ of a
counterion $i$ from a chain is defined as the distance of the
counterion from its closest monomer in space. We denote the set of
counterions belonging to one chain with ${\cal C}$.  From this we can
calculate $P(r)$ as

\begin{equation}
P(r) = \frac{1}{\QP} \int_{r^\prime=0}^r \mbox{d}r^\prime \sum_{i\in
  {\cal C}} \delta(r^\prime-d_i) \label{iondis}
\end{equation}

$P(r)$ denotes the fraction of counterions which are inside a shell with
radius $r$ around a polyelectrolyte chain. This definition can be used for a
large variety of chain conformations.

For an automated analysis of all types of pearl-necklace structures
appearing in our configurational data we need a tool to determine for
each monomer, to which pearl or string it belongs, and what is the
total number of such substructures.  First we should state that there
is no sharp definition of a pearl or a string. So what we need is a
practical approach to the problem. Our guideline in the development of
an automated tool for the identification of pearls and strings is that
the result should be close to the result which would be obtained by
looking at the conformation by eye.

Before we give a detailed explanation of the used cluster recognition
algorithm we want to present some other methods that are based on
observables which could be accessible with experimental methods. In
this way one can also judge how easily theses structures can be
observed by current experimental techniques.

\begin{figure}[ht]  \begin{center}   
    \includegraphics[angle=0,width=0.95\linewidth]{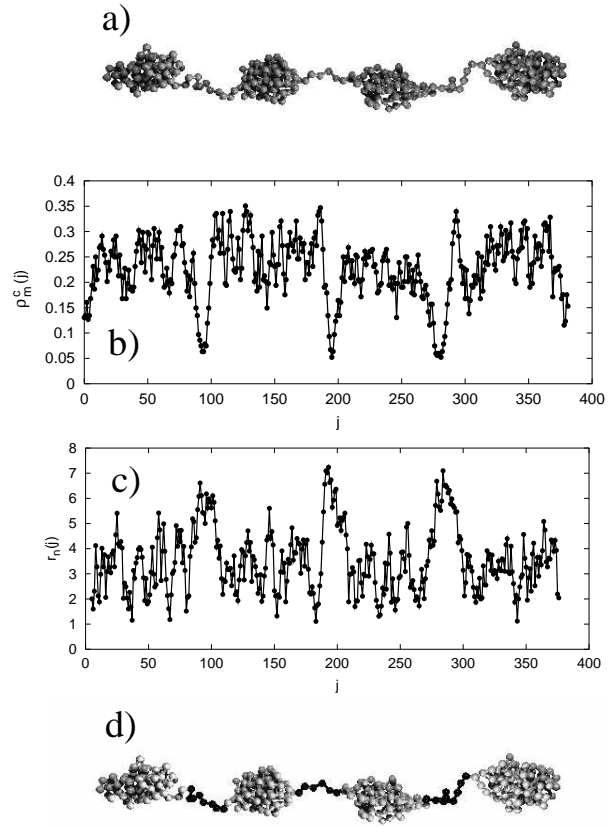}
    \caption{ 
      a) Snapshot of a polyelectrolyte chain with $\NM=382$ from
      series A1.  b) Coarse grained local monomer concentration
      $\rhom^c(j)$ with $r_c=4\sigma$ along the backbone.  c)
      Local distances $r_n(j)$ for $n=10$ along the backbone. d)
      Analysis with the cluster recognition algorithm. Monomers in
      pearls are grey and monomers in strings black.
      \label{fig_ana_trials}
    }
\end{center} \end{figure}

The local monomer concentration $\rhom(\V{r})$ of a chain is given by
$\rhom(\V{r}) = \sum_i^{\NM} \delta (\V{r} - \V{r}_i )$. To
distinguish the different substructures, pearls and strings, it is
better to use a coarse grained local monomer concentration
$\rhom^c(\V{r})$ which is defined as

\begin{equation}
\rhom^c(\V{r}) = \frac{1}{V_c} \, \int_{\V{r}^\prime \in V_c}
\mbox{d} \V{r}^\prime \, \rhom(\V{r}^\prime) \mbox{,}
\end{equation}

where $V_c$ is a spherical volume with radius $r_c$ around $\V{r}$.
This observable is larger in a pearl than in a string.  One can
calculate $\rhom^c(\V{r})$ along the backbone as a function of the
monomer positions $\V{r}_j$. This is done for a chain with length
$\NM=382$ from series A1 whose snapshot is shown in
fig.~\ref{fig_ana_trials}a. The resulting coarse grained local monomer
concentration with $r_c=4\sigma$ is shown for comparison directly
underneath in fig.~\ref{fig_ana_trials}b.

A second approach uses local distances between the monomers which
could be probed by NMR experiments~\cite{lee02a}. The local distance
between monomer $j$ and the monomer $j+n$ which is $n$ monomers apart
on the backbone is defined as

\begin{equation}
r_n(j) = \sqrt{(\V{r}_j - \V{r}_{j+n})^2} \mbox{ .}
\end{equation}

From scaling arguments one expects that the local distances inside the
compact pearls scale with $r_n \simeq n^{1/2}$ for small $n$ whereas
in the extended strings it scales with $r_n \simeq n$. The symbol
$\simeq$ is used to state a scaling relation which ignores numerical
prefactors. For the conformation shown in fig.~\ref{fig_ana_trials}a
the local distances along the backbone are shown for $n=10$ in
fig.~\ref{fig_ana_trials}c.

In both cases one can see the position of the 4 pearls and 3 strings.
The problem is that the variation of both $\rhom^c(\V{r})$ and
$r_n(j)$ is of the same order as the difference between the mean
values in pearls and strings.

Since the two previous methods have several problems with the
structure recognition of pearl-necklace conformations we have
developed a more reliable algorithm based on a simple cluster
recognition.  The question whether a group of monomers forms a cluster
is often connected to a distance criterion, e.g. monomers with a
distance smaller than a critical value belong to the same cluster.
For a polymer one has also to take into account the chain
connectivity.  This implies that a pure distance criterion is not
sufficient.  Therefore we require in addition that there are a certain
number of bonds between a pair of monomers along the chain contour.
The resulting algorithm is iterative and contains the following steps:

\begin{enumerate}
\item At the beginning every monomer is a cluster of size 1 (size =
  number of monomers belonging to a cluster).
\item Two clusters ${\cal C}1$ and ${\cal C}2$ are merged if they
  contain a pair of monomers $ij$ with $i \in {\cal C}1$, $j \in {\cal
    C}2$ and $r_{ij} < r_c$. In addition $i$ and $j$ are further apart
  than $n_c$ bonds along the chain contour: $|i-j|>n_c$.
\item Step 2 is repeated between all clusters as long as one finds
  clusters that have to be merged.
\item Remove loops: a cluster ${\cal C}1$ where all monomers are
  inside (along the chain contour) of another cluster ${\cal C}2$ are
  merged. Note that this step is only suitable for polyelectrolytes,
  but e.g. is not applicable for polyampholytes.
\item Practical definition of pearls: all clusters with a size larger
  than or equal $p_c$ are pearls. Pearls which are connected directly
  along the chain contour are merged.
\item Practical definition of strings: all clusters with a size
  smaller than $p_c$ belong to strings. Strings which are connected
  directly along the chain contour are merged.
\item Remove dangling ends and merge them to the end pearls. We do
  this because we have so far not seen dangling ends containing more
  than 3 monomers which could be seen as an extra string.
\end{enumerate}

Thus the algorithm has in principle three free parameters: $r_c$,
$n_c$ and $p_c$. But looking at the involved structures, one can
establish relations between these parameters which can be used as a
rough guide in choosing them. From scaling arguments we know that the
distance of two monomers $i$ and $j$ with a distance $|i-j|$ along the
chain contour scales as $|i-j|^{1/2}$ inside the pearls and with
$|i-j|$ inside the strings. Thus we can choose $b n_c^{1/2} < r_c < b
n_c$ to distinguish the two cases. For a weakly charged chain one can
determine a suitable value for $p_c$ with help of the pearl size
defined in eq.~\ref{gp_scaling}. But this is not possible for our
strongly charged systems with their subtle dependence on the
counterion distribution. The whole data analysis in this paper is done
with an empirical parameter set: $r_c= 2.1\sigma$, $n_c=6$ and
$p_c=9$.  Beside extended visual checks we have tested that small
changes of the three parameters do not have a significant effect on
the final result. The derived sizes for the substructures contain a
systematic error of $\pm4$ monomers coming from the two outer monomers
on each side of a substructure, for which one can not decide whether
they belong to the next pearl or the next string. A typical result for
a structure type with 4 pearls is shown in
fig.~\ref{fig_ana_trials}~d.  The four pearls and three strings
contain the following numbers of monomers: 90 - 8 - 94 - 6 - 77 - 9 -
98.  We do not claim that this is the fastest or the best way to
identify pearl-necklace structures, but it worked well when compared
with visual checks and thus served our purpose.  For an average pearl
size larger than 30 monomers it yields a reliability well above 95\%.

\section{Scaling}\label{sec:scaling}

One of our goals in this paper is to show to what extend scaling
theories that are made for long chains at infinite dilution can be
expected to work for dilute PE systems with finite length at finite
density.  The scaling theory predicts the dependencies of observables
like the chain extension on various parameters, e.g. $\NM$, $\lb$, $f$
and the reduced temperature $\tau_\mys{r}$. Here we give only a short
overview of some results of the scaling theory for the pearl-necklace
regime of polyelectrolytes. For the pearl-necklace regime one finds
the following relations~\cite{dobrynin96a,schiessel98a,lyulin99a}:
End-to-end distance $\RE$
\begin{equation}
\RE \simeq \NM b^{1/2} \lb^{1/2} f \tau_\mys{r}^{-1/2}, \label{pn_scaling1}
\end{equation}

number of pearls $\np$
\begin{equation}
\np \simeq \NM b^{-1} \lb f^{2} \tau_\mys{r}^{-1}, \label{np_scaling}
\end{equation}

string length $\ls$
\begin{equation}
\ls \simeq b^{3/2} \lb^{-1/2} f^{-1} \tau_\mys{r}^{1/2} , \label{ls_scaling}
\end{equation}

pearl size (radius) $\rp$
\begin{equation}
\rp \simeq b^{4/3} \lb^{-1/3} f^{-2/3},  \label{rp_scaling}
\end{equation}

pearl size (number of monomers) $\gp$
\begin{equation}
\gp \simeq b \lb^{-1} f^{-2} \tau_\mys{r} ,\label{gp_scaling}
\end{equation}

density inside the pearls $\rhop$:
\begin{equation}
\rhop \simeq b^{-3} \tau_\mys{r}. \label{rhop_scaling}
\end{equation}

Note that $\rp$ is independent of the solvent quality, and that our
solvent quality parameter $\elj$ is proportional to the second virial
coefficient of the LJ potential and hence also proportional to the
reduced temperature $\tau_\mys{r}$.

\subsection{Scaling variable chain length $\NM$}

\begin{figure}[ht]  
  \begin{center} 
    \includegraphics[angle=0,width=1.25\linewidth]{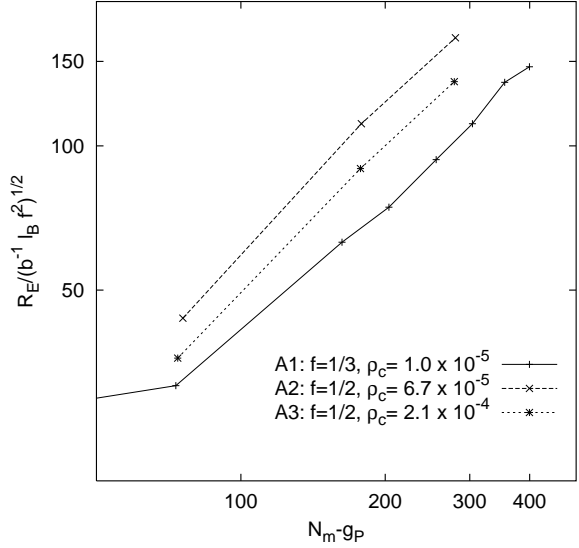}
    \caption{Scaling plot for $\RE$ versus $(\NM-\gp)$ (see
      eq.~\ref{pn_scaling1}). The series A1, A2 and A3 differ in the
      charge fraction $f$ and the density $\rhoc$. One can see that
      $\RE$ scales linearly with $(\NM-\gp)$, but the prefactor is not
      constant.
      \label{re_ndep}}
  \end{center} 
\end{figure}

The linear scaling of $\RE$ with $\NM$ is caused by the electrostatic
repulsion of the chain charge and is dominated by the string length
$\ls$.  Due to the finite length of our systems we have to correct the
scaling relation given in eq.~\ref{pn_scaling1}. For small chain
length, $\NM \le \gp$, the chain conformation consists of one pearl.
The size of this pearl scales as $\NM^{1/3}$.  Thus we have to replace
$\NM$ by $(\NM - \gp)$ in eq.~\ref{pn_scaling1}.  In
fig.~\ref{re_ndep} a linear relation between $\RE$ and $(\NM-\gp)$ can
be observed.
But the full scaling relation from eq.~\ref{pn_scaling1} including the
parameters specifying the Coulomb interaction, namely $f$, is not
valid.  Series A1 is performed at $f=\frac{1}{3}$ and its density is
in between that of series A2 and A3 which have both $f=\frac{1}{2}$.

\begin{figure}[ht]  
  \begin{center} 
    \includegraphics[angle=0,width=1.25\linewidth]{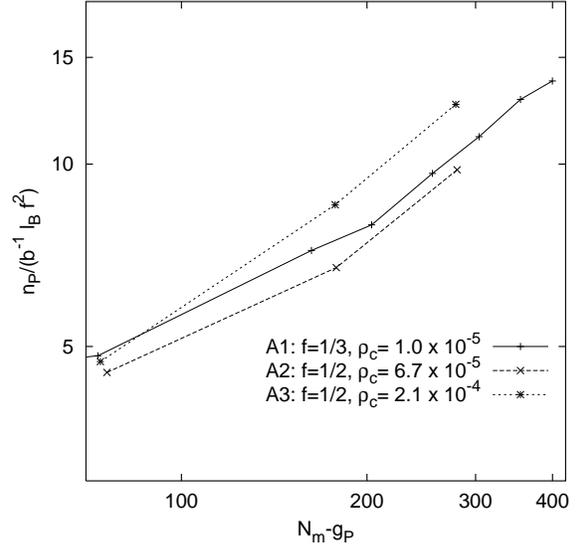}
    \caption{Scaling plot for $\np$ versus $(\NM-\gp)$ (see
      eq.~\ref{np_scaling}) for series A1, A2 and A3. \label{np_ndep}}
  \end{center} 
\end{figure}

We find a similar result for the number of pearls $\np$ in the
pearl-necklace regime. The scaling relation for $\np$ versus
$(\NM-\gp)$ (see eq.~\ref{np_scaling}) is shown for the data from
series A1, A2 and A3 in fig.~\ref{np_ndep}.
Again there is a linear relation between $\np$ and $\NM$ but different
prefactors for different values of $f$ and $\rhoc$.

\begin{figure}[ht]  \begin{center}  
    \includegraphics[angle=0,width=0.95\linewidth]{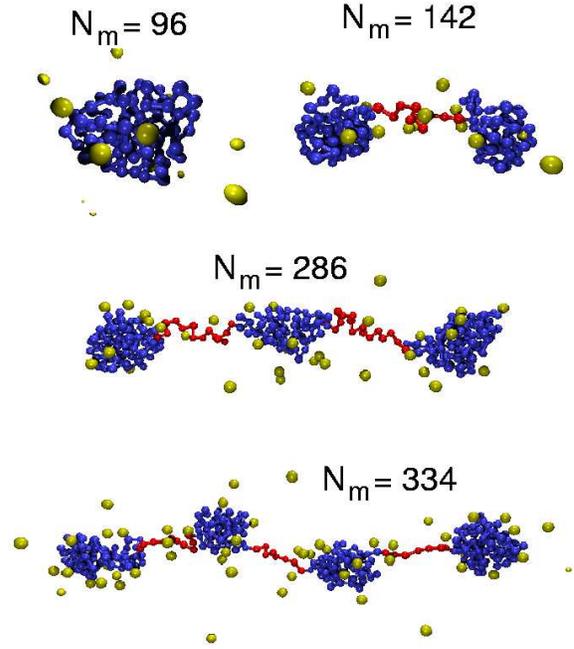}
    \caption{Snapshots from series A1 with different chain length $\NM$ and
      different number of pearls $\np$.
      \label{phase_spaceN}}
\end{center} \end{figure}

To get some visual impression, we show in fig.~\ref{phase_spaceN} some
snapshots of simulations with different chain length $\NM$ from series
A1.
The discreteness of the pearl number $\np$ does not play a significant
role for sufficiently long chains which give rise at least to a
dumbbell. Then our data is in accord with the scaling relations for
the quantities pearl size, string length and pearl density, namely
they are constant within the statistical error. We find for the
systems of series A1 with $\NM>200$: $\gp = 78 \pm 4$, $\rhop = 0.67
\pm 0.04\sigma^{-3} $ and $\ls = 7.3 \pm 2 \sigma$.  As we will show
later, the discreteness of the number of pearls is smeared out due to
fluctuations between different structure types in a way that the
average quantities can maintain their optimal values, compare also
section~\ref{sec_fluctuations}.

\subsection{Solvent quality $\elj$ }\label{subseq:solvent}

In the simulation series B we have tested the behavior of
polyelectrolyte chains upon changing the solvent quality via the
Lennard-Jones parameter $\elj$.  Since the practically usable range
for this parameter is small, ranging from $\elj=0\KB T$ to $2.0\KB T$,
it is not possible to test the scaling predictions over a large
parameter range. In addition the range of $\elj$ values for which we
observed pearl-necklace structures is even smaller, namely from $\elj
= 1.0 \KB T$ to $2.0\KB T$.  For values of $\elj < 1.0 \KB T$ one
leaves the poor solvent regime and for values $\elj > 2 \KB T$ we
encounter simulational problems, e.g. kinetically frozen states.

\begin{figure}[ht]  
  \begin{center}   
    \includegraphics[angle=0,width=0.95\linewidth]{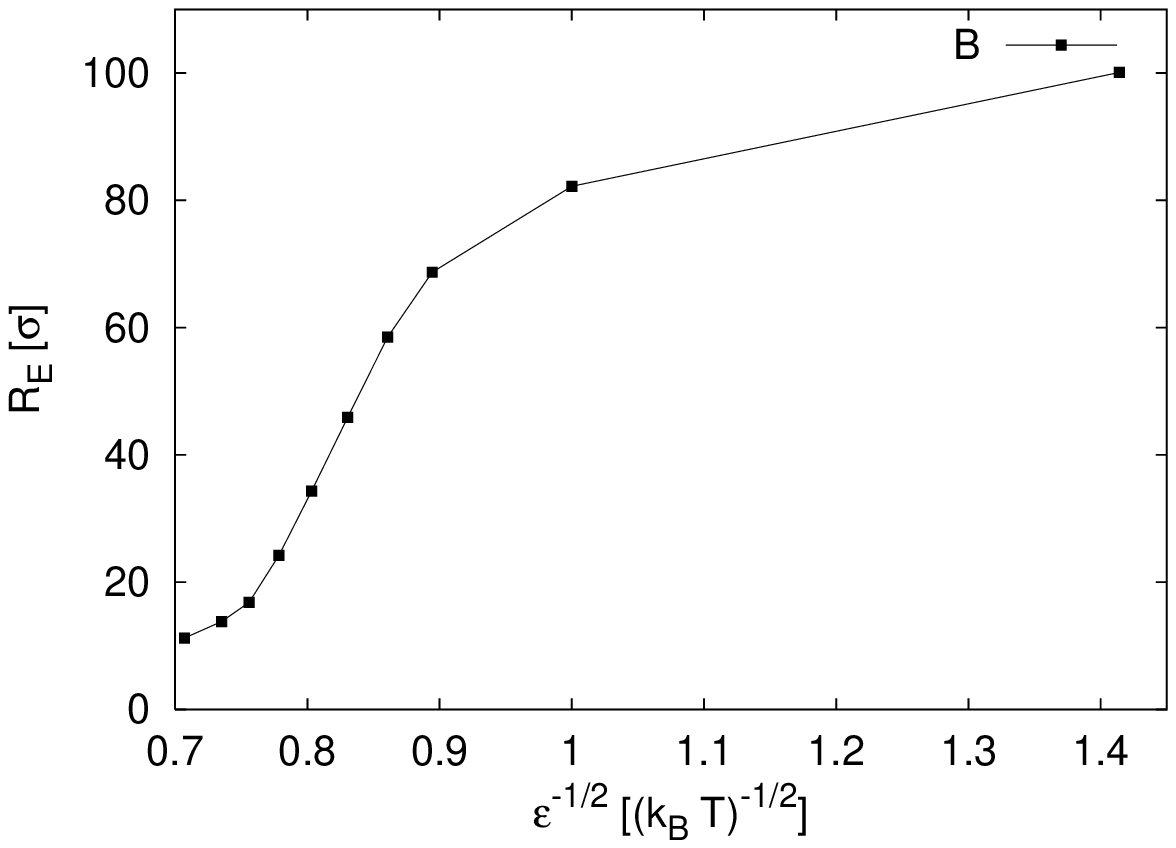}
    \caption{Dependence of $\RE$ on the solvent
      quality parameter $\elj$ (simulation series B).
      \label{re_elj_dep}}
  \end{center} 
\end{figure}

\begin{figure}[ht]  
  \begin{center}   
    \includegraphics[angle=0,width=0.95\linewidth]{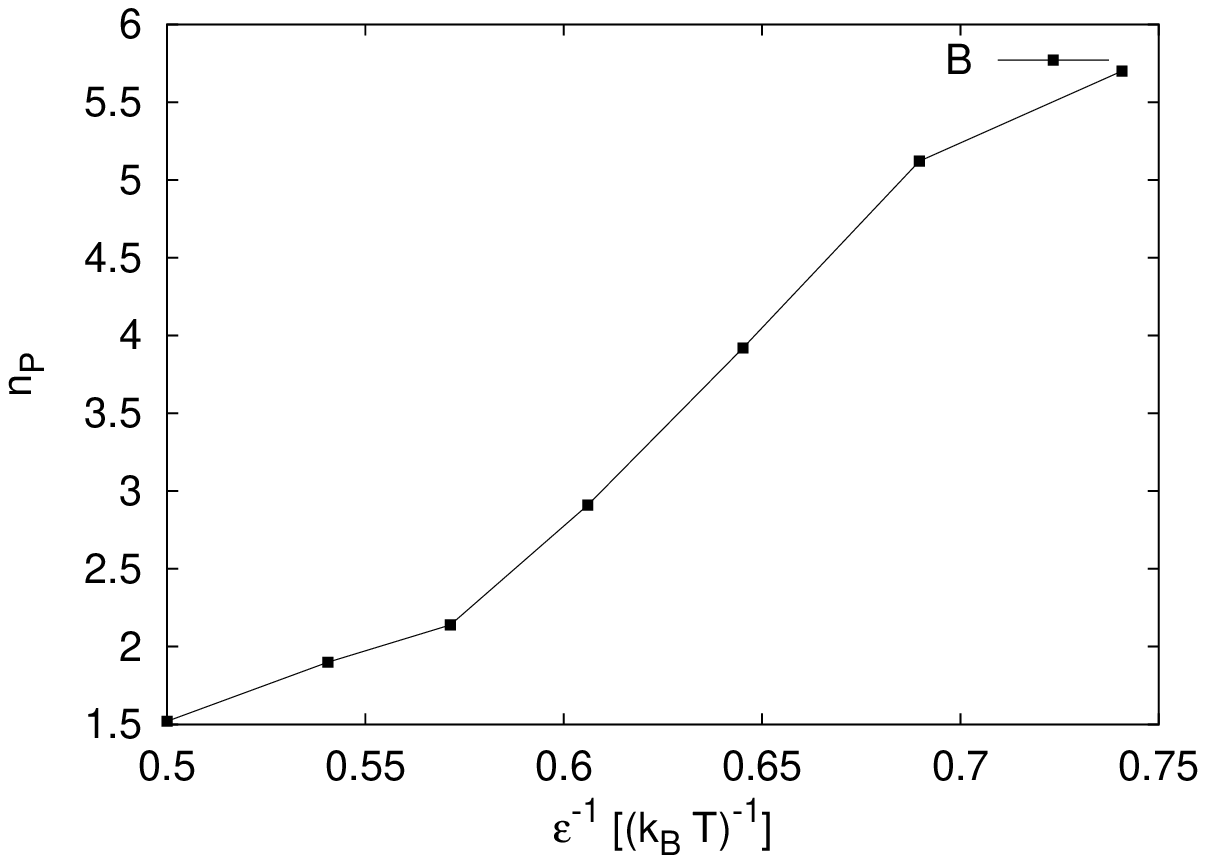}
    \caption{Dependence of $\np$ on the solvent
      quality parameter $\elj$ (simulation series B).
      \label{np_elj_dep}}
  \end{center} 
\end{figure}

\begin{figure}[ht]  
  \begin{center}   
    \includegraphics[angle=0,width=0.95\linewidth]{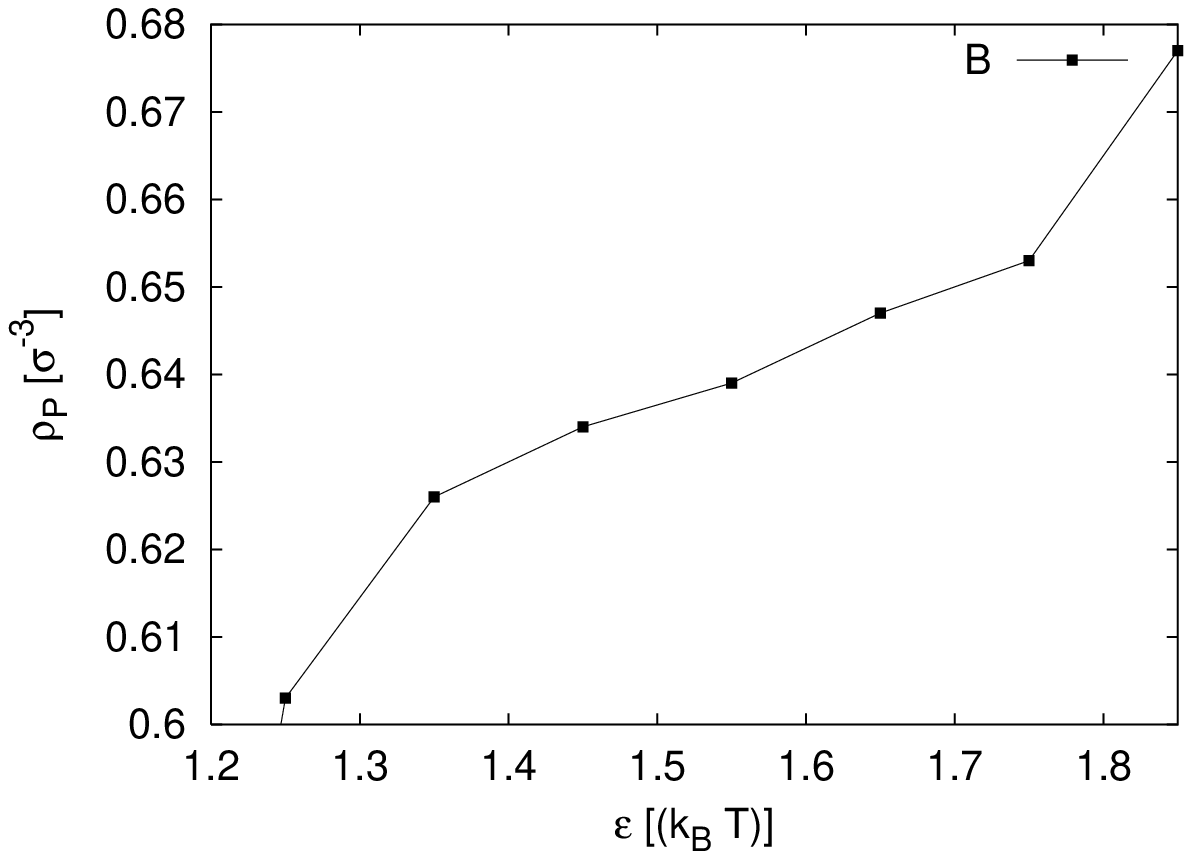}
    \caption{Dependence of $\rhop$ on the solvent quality parameter
      $\elj$ (simulation series B). \label{rhop_elj_dep}}
  \end{center} 
\end{figure}

\begin{figure}[ht]  
  \begin{center}   
    \includegraphics[angle=0,width=0.95\linewidth]{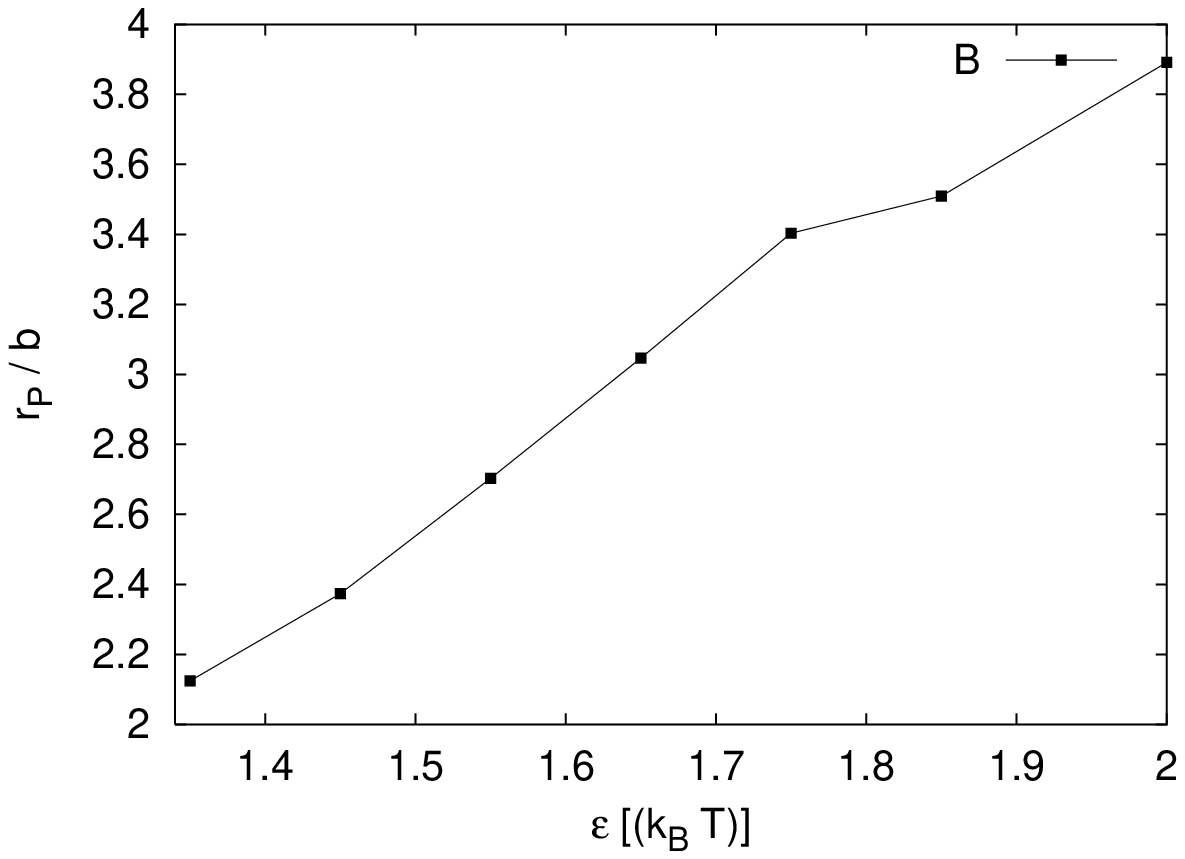}
    \caption{Dependence of $\rp$ on the solvent quality parameter
      $\elj$ (simulation series B). \label{rp_elj_dep}}
  \end{center} 
\end{figure}

The dependence of $\RE$ and $\np$ on $\elj$ is shown in
fig.~\ref{re_elj_dep} and fig.~\ref{np_elj_dep}. 
For both observables the scaling predictions seem to hold for a small
$\elj$-region. $\RE$ scales linear with $\elj^{-1/2}$ in the range
from $\elj=1.25\KB T$ to $\elj=1.75\KB T$ (see eq.~\ref{pn_scaling1}).
The regime where $\np$ scales as $\elj^{-1}$ is even smaller and
extends from $\elj=1.45\KB T$ to $\elj=1.75\KB T$ (see
eq.~\ref{np_scaling}).  In the same regime also the density inside the
pearls $\rhop$ which is plotted against $\elj$ in
fig.~\ref{rhop_elj_dep} shows a linear dependence on $\elj$ as
predicted by eq.~\ref{rhop_scaling}.
However, a more reliable conclusion may be drawn from the scaling
relation for the pearl size $\rp$, since $\rp$ should be independent
of the solvent quality (see eq.~\ref{rp_scaling}). We therefore have
calculated the pearl size as $\rp = \frac{3}{4}\pi \gp \rhop^{-1/3}$
from our simulation data and plotted it versus $\elj$. As can be seen
in fig.~\ref{rp_elj_dep} $\rp$ increases monotonically with $\elj$,
hence shows an unexpected dependency on $\elj$.
We are lead therefore to conclude that the scaling predictions for the
dependency on the solvent quality do not work here. They fail for the
most definite test case, the independence of the pearl radius $\rp$ of
the solvent quality. Nevertheless in a small regime for $\elj /(\KB T)
= 1.45 \dots 1.75$ they seem to work approximately for some
observables due to a fortuitous error cancellation.

\begin{figure}[ht]  
  \begin{center}   
    \includegraphics[angle=0,width=0.95\linewidth]{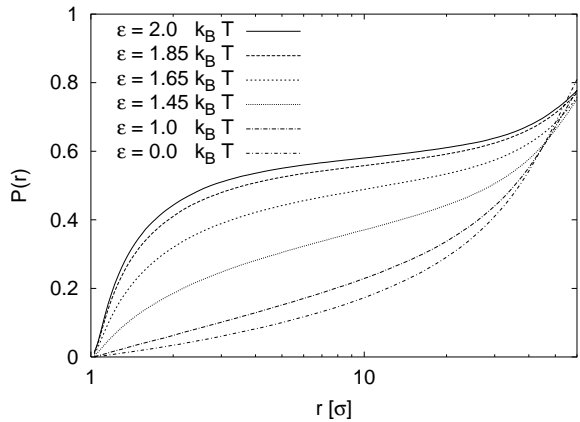}
    \caption{Integrated counterion distribution $P(r)$ around
      polyelectrolyte chains subject to different solvent qualities.
      \label{iondis_eljdep}}
  \end{center} 
\end{figure}

To elucidate the role of the finite counterion density around the
chains we calculated the integrated ion distribution $P(r)$ (see
eq.~\ref{iondis}).  In fig.~\ref{iondis_eljdep} $P(r)$ is shown for
different solvent qualities.
The counterion distribution changes greatly close to the chain with
$\elj$.  In the pearl-necklace regime between $\elj/(\KBT) =1.0$ and
$\elj/(\KBT)=1.85$ which is the same regime where we find also the
strongest change in the chain extension we observe that the counterion
distribution is changing the most.  The fraction of counterions being
very close to the chain which hence can be called condensed is also
varying strongly. Only the top four curves show an inflection point
which is a sign of counterion
condensation~\cite{belloni98a,deserno00a} (see also
sec.~\ref{coexistance}).  Scaling theory assumes usually that the
counterion condensation depends only on the Manning parameter $\xi =
\lb b^{-1} f$ and is hence supposed to be independent of the actual
conformation of the chain.  Note that the total series B is performed
at $\xi=0.5$ and is thus expected to show no counterion condensation
at all according to the standard
Manning-Oosawa~\cite{manning69a,oosawa71a} concept which requires $\xi
\ge 1$.

\subsection{The Coulomb parameters: $\lb$ and $f$}

\begin{figure}[ht]  
  \begin{center}   
    \includegraphics[angle=0,width=0.95\linewidth]{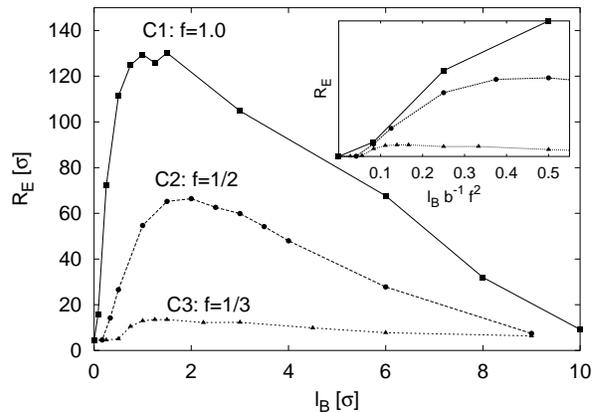}
    \caption{Dependence of $\RE$ on the Bjerrum length $\lb$ for the
      series C1, C2 and C3. The inlay shows the dependence of
      $\RE$ on the scaling variable $\lb b^{-1} f^2$.
      \label{fig.re_lbdep}}
  \end{center} 
\end{figure}

\begin{figure}[ht]  
  \begin{center}   
    \includegraphics[angle=0,width=0.95\linewidth]{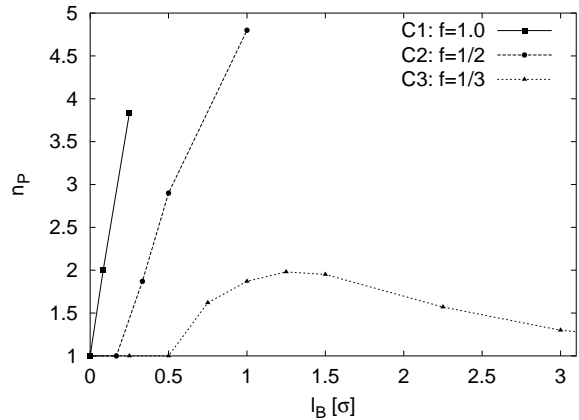}
    \caption{Dependence of $\np$ on the Bjerrum length $\lb$ for the
      series C1, C2 and C3.
      \label{fig.np_lbdep}}
  \end{center} 
\end{figure}

As we demonstrated in the previous subsection~\ref{subseq:solvent},
the interplay between counterions and chain conformation influences
greatly the behavior of PEs in poor solvent. Therefore also a thorough
investigation of the parameters determining the Coulomb interactions
is necessary.  This section treats the simulation series C1, C2 and C3
where we have investigated the $\lb$ dependence but using different
charge fractions $f=1$(C1), $f=1/2$(C2) and $f=1/3$(C3).  All
simulations are performed at the same charge density $\rhoc$.  We
remark that the valences of the charged particles is yet another
important independent parameter. But since this is also not included
in the scaling picture and would even further complicate the picture
we will leave this for another study.  We will also look at the
behavior of these systems with regard to parameters combined of $\lb$
and $f$, namely the Manning parameter $\xi$ and the scaling variable
$\lb b^{-1} f^2$. The Manning parameter is important for the
electrostatic field that counterions would experience around a long
stretched chain and its value determines in a first approximation the
onset of counterion condensation.  In the framework of scaling theory
simulations performed at the same value of $\lb b^{-1} f^2$ should be
identical.  Nevertheless our simulations show a big difference between
the three series. The dependence of $\RE$ on $\lb$ is shown in
fig.~\ref{fig.re_lbdep} and that on $\np$ in fig.~\ref{fig.np_lbdep}.
Only for very small values of $\lb$ the counterions and their
interaction with the chain conformation do not play a significant
role. The prediction from scaling theory (see eq.~\ref{pn_scaling1})
is that our data should collapse on a single master curve. However,
our data show that this is only true if the scaling variable $\lb
b^{-1} f^2$ is smaller than $0.1$. This can be seen in the inset in
the upper right corner of fig.~\ref{fig.re_lbdep} where we plotted the
same data versus $\lb b^{-1} f^2$. Upon a further increase of $\lb
b^{-1} f^2$ the values of the end-to-end distances diverge rapidly.
Whereas the chains of series C3 ($f=1/3$) already start to shrink at
$\lb b^{-1} f^2 = 0.15$ the chains of the other series still expand.

\begin{figure}[ht]  
  \begin{center}   
    \includegraphics[angle=0,width=0.95\linewidth]{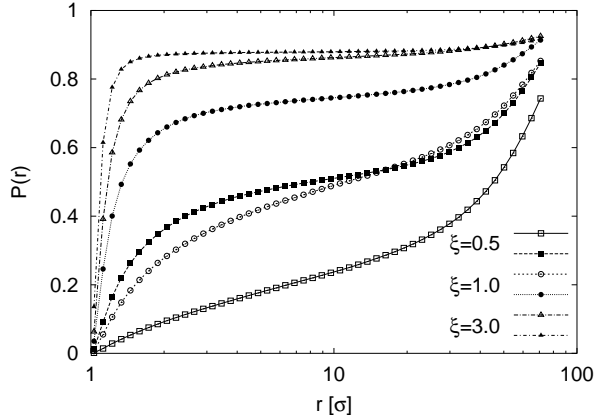}
    \caption{Integrated counterion distribution $P(r)$ around
      polyelectrolyte chains for different $\xi$-values. Open symbols
      show data from series C2 with $f=1/2$ and filled symbols from C3
      with $f=1/3$.
      \label{iondis_xidep}}
  \end{center} 
\end{figure}

The different values of $f$ can be seen as different schemes for the
discretization of the backbone charge. This has an effect on the
correlations between the charges which influences the chain
conformation and thus also the counterion distribution. In
fig.~\ref{iondis_xidep} we have therefore plotted the integrated
counterion distribution $P(r)$ for several values of the Manning
parameter $\xi$ for series C2 and C3.
Already the curves for $\xi = 0.5$ which correspond to $\lb=1\sigma$
for series C2 and $\lb=1.5\sigma$ for series C3 show a pronounced
difference in $P(r)$. Even though the effect of $f$ on the charge
charge correlations may be small it is enhanced strongly by the
interplay between the chain conformation and the counterion
distribution.  When the counterions only slightly move towards the
chain due to correlation effects, the effective charge of the chain
will shrink and so does the end-to-end distance. A higher effective
Manning parameter $\xi_{\mys{$\RE$}} := \QP b / \RE$ follows which
again attracts more counterions towards the chain. The same holds for
the opposite way. This is the same mechanism which we have already
seen for the dependence on $\elj$ in the previous section. The
difference in $P(r)$ becomes smaller with increasing $\xi$ which is
due to the gradual collapse of the polyelectrolyte chain, since then
in all cases most of the counterions are close or even inside the
chains.

\begin{figure}[ht]  
  \begin{center}   
    \includegraphics[angle=0,width=0.95\linewidth]{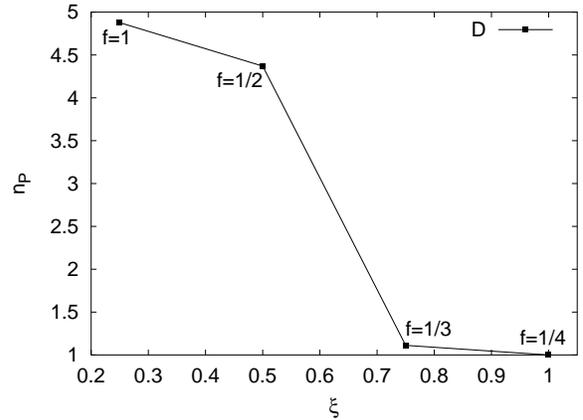}
    \caption{Dependence of $\np$ on the Manning parameter $\xi$ for
      the simulation series D with constant scaling parameter $\lb
      b^{-1} f^2 = 0.25$. \label{np_lbf2}}

  \end{center} 
\end{figure}

\begin{figure}[ht] 
  \begin{center} 
    \includegraphics[angle=0,width=0.95\linewidth]{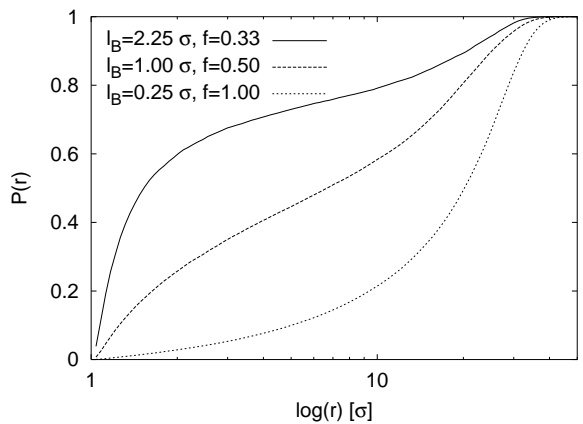}
    \caption{Integrated counterion distributions $P(r)$ are shown for
      different charge fractions $f$ with constant $\lb b^{-1}
      f^2=0.25$ (series D). \label{ps_id_lbf}}
  \end{center} 
\end{figure}

In series D we have performed four simulations with $\lb b^{-1}
f^2=0.25$, but different values for $\lb$ and $f$.  In
fig.~\ref{np_lbf2} we show the behavior of $\np$ as a function of
$\xi$.
Decreasing $f$ and thus increasing $\lb$ has again a drastic effect on
the chain conformation leading to collapse of the chains for large
$\lb$ and small $f$. Even though the number of pearls for the two
first points with $f=1$ and $f=1/2$ do not differ much the chain
extension shows a large difference, namely $\RE=59\sigma$ for $f=1$
and $\RE=32\sigma$ for $f=1/2$. This is also reflected in a large
difference of the pearl sizes and the string lengths. The difference
between the systems can again be traced to a quite different
distribution of the counterions as can be inspected in
fig.~\ref{ps_id_lbf} where we show $P(r)$ for three of the simulations
of series D.
It is really surprising that such a small change, basically in the
discretization of the charges, has such a big impact on the chain
conformations. This again shows that the delicate balance between
repulsive and attractive forces is very sensitive to subtle
changes. These effects are definitely not captured by the parameter
$\lb b^{-1} f^2$ which is used in scaling theories. We will come back
to the collapse discussion in sec.~\ref{sec_pase_space} where we will
also give an overview of the occurring conformations. 

\section{Fluctuations \label{sec_fluctuations}}

The data analysis shows that pearl-necklace conformations are very
soft objects which display large fluctuations on all length scales. We
will measure in detail the extent of the various fluctuations, since
this is important for the interpretation of experimental measurements
on polyelectrolyte solutions (see section~\ref{sec_formfac}).

\subsection{Fluctuations of the structure type - Coexistence \label{coexistance}}

\begin{figure}[ht]  
  \begin{center}   
    \includegraphics[angle=0,width=0.95\linewidth]{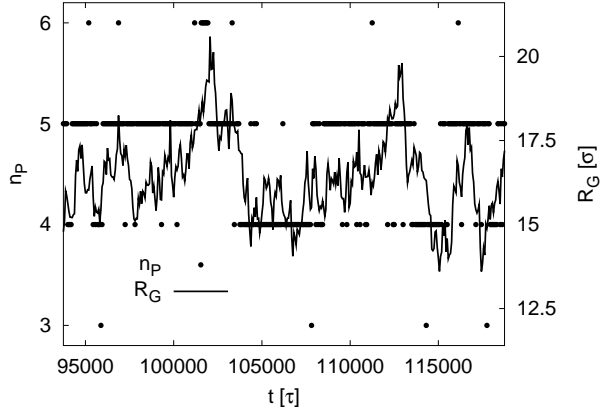}
    \caption{Time evolution of the structure type (number of pearls)
      and the radius of gyration $\RG$ of a single chain ($\NM=382$,
      series A1).
      \label{struc_dev}}
  \end{center} 
\end{figure}

In most of our simulated systems we find coexistence of pear-necklaces
with different number of pearls which we call structure types.  It is
important to note that the observed coexistence is not caused by
frozen meta-stable states. We have excluded this possibility by
observing individual chains over a larger period of time each showing
a large number of transitions between different structure types during
the simulation time. The typical time evolution of the structure type
$\np$ and the radius of gyration $\RG$ for an individual chain is
exemplarily shown in fig.~\ref{struc_dev}.
In this case the chain mainly fluctuates between structures with 4 and
5 pearls, but we also see a significant fraction of structures with 3
and 6 pearls. Looking at $\RG$ one can see a certain correlation with
the structure type which is reflected also in the mean value for
different structure types: $\RG^{(3)} = (14.7\pm1.3) \sigma$,
$\RG^{(4)} = (16.1\pm1.5) \sigma$, $\RG^{(5)} = (17.5\pm1.6) \sigma$,
$\RG^{(6)} = (19.5\pm1.7) \sigma$. Here and in the following the
superscript denotes that a observable is measured only for
conformations with a certain number of pearls $\np$. But nevertheless
such a simple chain observable is not suited for structural
discrimination as one can see from the given mean deviations. The same
holds for other chain observables like $\RH$, the characteristic
ratios $r$ and $\alpha$.

\begin{figure}[ht]  
  \begin{center}   
    \includegraphics[angle=0,width=0.95\linewidth]{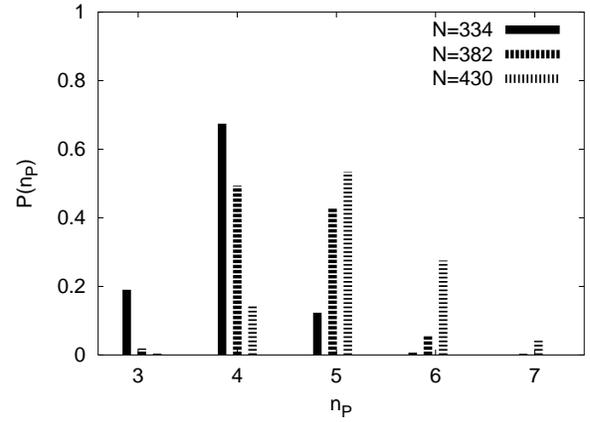}
    \caption{Probability distribution for the structure types found in 
      systems with different chain length from series A1.
      \label{structype_dist}}
  \end{center}
\end{figure}

One could be tempted to explain the coexistence of two structure types
by a simple finite size argument, namely that the ratio of $\NM$ and
the optimal pearl size $\gp$ is not always an integer. In
fig.~\ref{structype_dist} we plot the probability distributions
$P(\np)$ of the structure types for three different chain lengths.
We clearly observe a coexistence range containing up to four different
structure types. This means that the different observed structure
types can only have small differences in the free energy of the order
of the thermal fluctuation spectrum. To confirm this we calculate from
the shown probability distribution the free energy differences between
structure types with $n$ and $m$ pearls according to the Boltzmann
factor:

\begin{equation} 
  \frac{\Delta {\cal F}^{(nm)}}{\KBT} = \frac{{\cal F}^{(n)} - {\cal
      F}^{(m)}}{\KBT} = \ln \frac{p^{(n)}}{p^{(m)}}
  \label{boltzmann} 
\end{equation} 

For the case with $\NM=430$ we find: $\Delta {\cal F}^{(45)} =
-1.32\,\KBT$, $\Delta {\cal F}^{(56)} = 0.66\,\KBT$, $\Delta {\cal
  F}^{(67)} = 1.9\,\KBT$. All values are of the order of $\KBT$ which
is consistent with our observed large coexistence regime.

\begin{figure}[ht]
  \begin{center}   
    \includegraphics[width=0.95\linewidth]{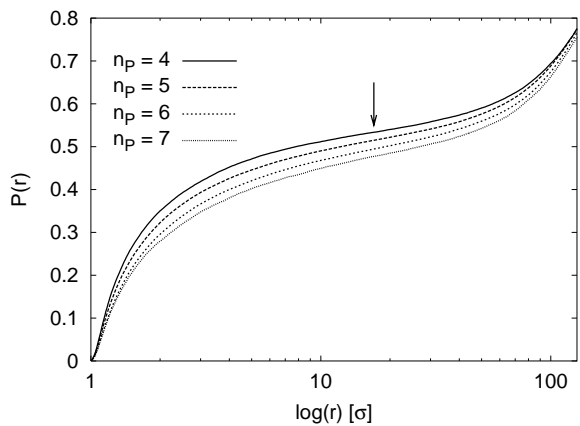}
    \caption{Integrated  counterion distributions $P(r)$ for different
      structure types for the system with $\NM=430$ from series A1. The
      arrow pointing to $(17 \sigma,\;0.53)$ shows the position of the
      inflection point for the structure type with $\np=4$.
      \label{id_strucdep} }
  \end{center}
\end{figure}

As we will argue these small differences in the free energy between
different structure types are mainly due to the interplay of the
counterion distribution and the chain conformation.  To elucidate the
role of the counterions we look at the counterion distribution around
the PEs analyzed for each structure type separately.  In
fig.~\ref{id_strucdep} we have plotted the integrated counterion
distributions $P(r)$ for different structure types for the system with
$\NM=430$ from series A1.
One observes more counterions to be close to the chains with a smaller
pearl number.  This can be easily understood if one looks at the far
electrostatic field of the chain. For distances larger than the
pearl-pearl separation $\rpp$ the chain can be seen as a charged
cylinder with an effective Manning charge parameter $\xi_{\mys{$\RE$}}
:= \QP b / \RE$. The end-to-end distance increases with increasing
pearl number: $\RE^{(4)}=48.8\sigma$, $\RE^{(5)}=54.1\sigma$,
$\RE^{(6)}=60.1\sigma$ and $\RE^{(7)}=64.0\sigma$, hence the shorter
chains have a larger effective Manning charge parameter, namely
$\xi_{\mys{$\RE$}}^{(4)} = 2.95$, $\xi_{\mys{$\RE$}}^{(5)} = 2.67$,
$\xi_{\mys{$\RE$}}^{(6)} = 2.40$ and $\xi_{\mys{$\RE$}}^{(7)} = 2.25$.

Since the PEs in this regime are elongated structures and they carry a
large effective line charge it might be worthwhile to compare the ion
distribution with predictions from Poisson-Boltzmann (PB) theory. This
is also supported by the functional form of the integrated counterion
distributions in fig.~\ref{id_strucdep} which looks very similar to
that of an infinitely charged rod within the cell
model~\cite{deserno00b}.  In the framework of the PB theory applied to
the cell model for an infinitely long charged
cylinder~\cite{deserno01c} one can calculate the fraction $f_c$ of
Manning-Oosawa condensed
counterions~\cite{deserno00b,manning69a,oosawa71a} as $f_c = 1 -
\frac{1}{\xi}$.  In the simulation the fraction of condensed
counterions can be read of as the value of $P(r)$ where the $P(r)$
curve, plotted as a function of $\log(r)$, has an inflection point.
Using $\xi=\xi_{\mys{$\RE$}}$ we find for the $\np = 4$ structure that
the cell model prediction for $f_c = 0.66$ is higher than the value
which can be read off the inflection point criterion, that is $f_c =
0.53$.  A perfect quantitative agreement can however not be expected
since our system is finite, and has as such a smaller electric field,
and secondly the pearl-necklace structures do alter the near
electrostatic field of the chain and lead to an inhomogeneous
counterion distribution. As a word of caution we remark that the bare
value for the Manning parameter of the chain in Fig.~\ref{id_strucdep}
is 0.5 which would not lead to any counterion condensation at all in
the simple Manning picture, and the functional form of an infinite rod
with that charge value would simply not show any inflection point.
This again demonstrates that poor solvent chains, due to the short
range attraction between monomers, feature a higher effective charge
density. Since the Poisson-Boltzmann cell model can at least
qualitatively explain the change in the counterion distributions a
more refined version of this model seems to be necessary. A possible
starting point could be to combine the PB rod and sphere geometry as
has been suggested in ref.~\cite{deshkovski01a}.

When more counterions are near the chain we find in terms of a charge
renormalization that the effective line charge density of the chain
decreases. This allows the chain to contract further which itself
induces a stronger counterion attraction. Again a lower effective line
charge density also increases the optimal pearl size. This results in
shorter chains with fewer pearls. The opposite, longer chains with
more pearls, holds when the counterions move away from the chain. This
lowers the difference in the free energy between the different
structure types, as was suggested above.

Scaling theories have predicted a collapse of the pearl-necklace
structure into a globule as soon as counterion condensation
starts~\cite{khokhlov80a,dobrynin96a,schiessel98a,vilgis00a} due to an
avalanche behavior of condensing counterions that contracts the chain.
In the investigated parameter range where we observe stable
pearl-necklace conformations in our simulation data we did not see
such a collapse transition.  In the light of a recent
study~\cite{deserno01f} this is not surprising, since it can be shown
that this collapse depends on how easily the counterions can enter the
globule, the ion concentration, and the strength of the electrostatic
interaction. Also the necessary amount of condensed counterions to
induce the collapse can be quite high.

We also suggest that the fluctuations due to the presence of the
counterions lower the energy barrier between the different structure
types. To confirm this suggestion one would need to analyze the
transition frequency between the different structure types for chains
with explicit counterions and a corresponding Debye-H\"uckel chain
with the same $\np$. This is however outside the scope of the present
investigations.

\subsection{Fluctuations of the substructures}

\begin{figure}[ht]  
  \begin{center} 
    \includegraphics[angle=0,width=0.95\linewidth]{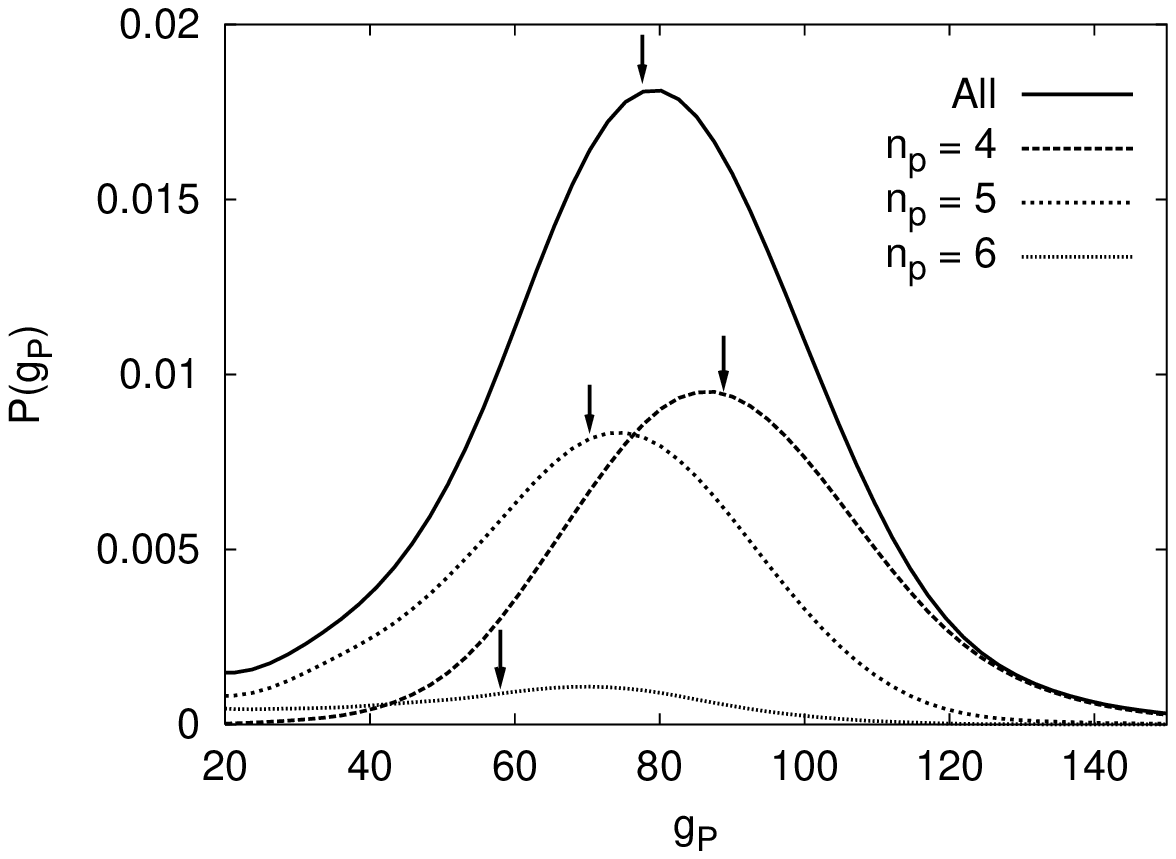}
    \caption{Probability distributions $P$ for the pearl size $\gp$
      for the system with chain length $\NM=382$ from series A1. Shown
      is the distribution for all chains as well as the distributions
      for the different structure types. The arrows mark the mean
      value of the corresponding probability distribution.
      \label{ps_ppd_dis}}
\end{center} \end{figure}

\begin{figure}[ht]  
  \begin{center} 
    \includegraphics[angle=0,width=0.95\linewidth]{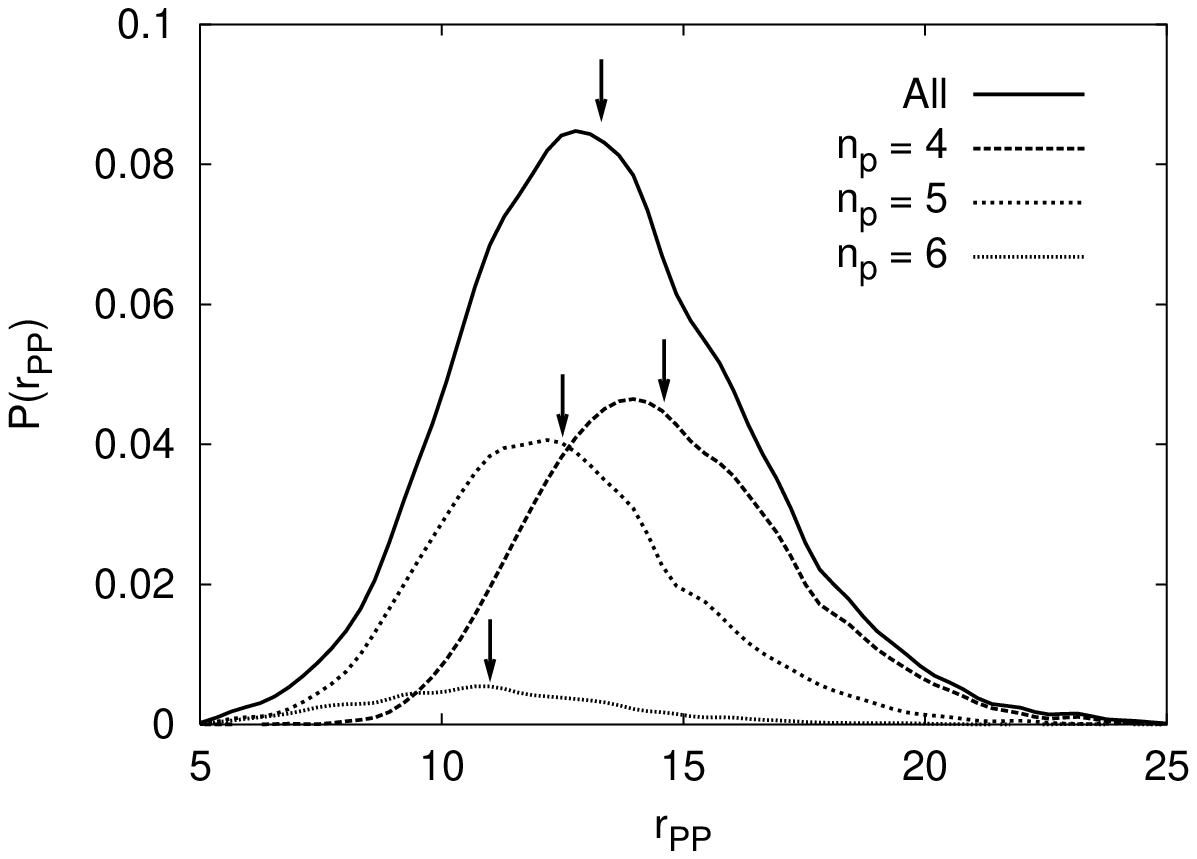}
    \caption{Probability distributions $P$ for the pearl-pearl
      distance $\rpp$ for the system with chain length $\NM=382$ from
      series A1. Shown is the distribution for all chains as well as
      the distributions for the different structure types. The arrows
      mark the mean value of the corresponding probability
      distribution.
      \label{ppd_dis}}
\end{center} \end{figure}

In this section we discuss fluctuations of the chain conformation on a
smaller length scale, namely the radius of the pearls $\rp$ and the
pearl-pearl distance $\rpp$.  The probability distributions for the
pearl size $\gp$ is shown in fig.~\ref{ps_ppd_dis} and for the
pearl-pearl distance $\rpp$ in fig.~\ref{ppd_dis}.
Both distributions are quite broad. The relative standard deviations
for the distributions of all chains are $\delta \gp = 0.32$ and
$\delta \rpp = 0.22$.

As we have already seen for the overall chain extension also the size
of the substructures are influenced by the counterion distribution.
The pearl size is decreasing with increasing number of pearls and
increasing chain extension as one can see from the arrows in
fig.~\ref{ps_ppd_dis}.  Still an explanation in terms of a charge
renormalization is not sufficient for an understanding. This becomes
more clear when we look at the number of counterions inside a shell
around the chains. Rounded to integers we find for a shell radius of
$3\sigma$ on average $56$ counterions. Averaged separately for the
different structure types we find $60$, $57$, $53$ and $50$
counterions for structures with $4$, $5$, $6$ and $7$ pearls. It is
interesting to note, that the difference in this number of counterions
is roughly constant for shells with radii ranging from $2\sigma$ up to
$30\sigma$ and is thus independent of the exact definition of
counterions called ``condensed'' to the chains. Thus we now use a
practical definition of an effective charge using the counterions
which are closer than $3\sigma$ to the next monomer. For the example
this yields an effective pearl charge of $18$, $15$, $13$ and $11$
charges for structures with $4$, $5$, $6$ and $7$ pearls respectively.
Since the pearl size is only slowly varying one gets large differences
in the electrostatic self energy of the pearls of different structure
types.

In the scaling Ansatz the distance between neighboring pearls $\rpp$
is determined by the balance between the electrostatic repulsion
between the chains and the energy one needs to pull monomers out of
the pearls. The second term is related to the surface tension and is
roughly constant.  The term connected to electrostatics is more
complex. We approximate it by the electrostatic energy of two point
charges carrying the average pearl charge, separated by the average
pearl-pearl distance.  $\rpp$ ranges from $15.4\sigma$ for $4$-pearl
structures to $11.1\sigma$ for $7$-pearl structures. Together with the
above calculated effective pearl charges we can estimate the
electrostatic repulsion between neighboring pearls to be $31.9\KBT$
for $4$-pearl structures and $16.7\KBT$ for $7$-pearl structures. Note
that we neglect the energetic contribution of the other pearls. This
difference can also not be explained by a standard Debye-H\"uckel
potential since the Debye screening length, calculated from the bulk
charge density $\rho_c=1\times10^{-5}\sigma^{-3}$ in the system, is
$\lambda_{D}=72.8\sigma$ and is thus much larger than the values for
$\rpp$. Of course, closer to the chain the counterion density is much
larger than in the bulk which could be a hint to explain the observed
differences. The counterion density is rapidly varying with the
distance from the chain starting with $\approx
\times10^{-1}\sigma^{-3}$ at a distance of $1.5\sigma$ and dropping
below the bulk density at a distance of $50\sigma$.

The electrostatic self energy of the pearls and the electrostatic
pearl-pearl repulsion show explicitly the discrepancies between a mean
field approach and our simulational results.  It also clearly
demonstrates that it is necessary to include the counterion
distribution into the model. Moreover correlations between the
counterions and the chain charges seem to play an important role. For
instance, it can be shown that the counterions preferentially
accumulate between the pearls.  Since the structure of
polyelectrolytes in poor solvent turns out to be extremely sensitive
to the inhomogeneities of the counterion distribution it should be
used as a test case for the development of theoretical approaches
beyond the mean field level. A more detailed discussion of these
inhomogeneities can be found in ref.~\cite{limbach01a,castelnovo00a}

\section{The sausage regime and phase space \label{sec_pase_space}}

\begin{figure}[ht]  \begin{center}  
    \includegraphics[angle=0,width=0.95\linewidth]{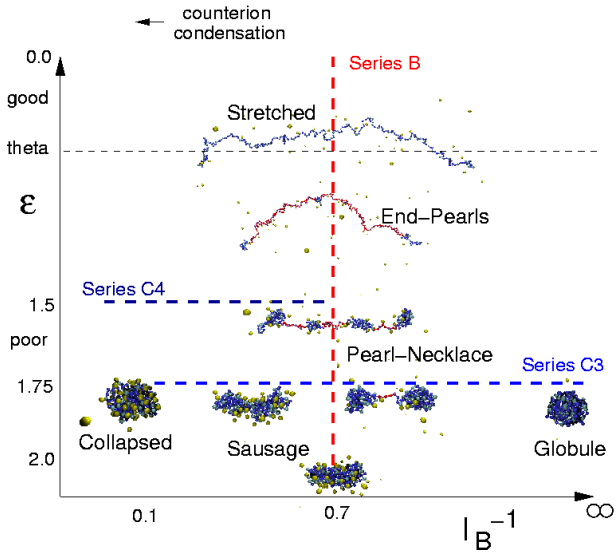}
    \caption{Schematic view of the location of the different PE
      configuration types in the $\elj$/$\lb^{-1}$ phase diagram. All
      simulations are performed at $f=\frac{1}{3}$. The dashed lines
      indicate the location of the simulation series B, C3 and C4. 
      \label{phase_space1}}
\end{center} \end{figure}

In order to get an overview of the parameter regimes our different
simulation series are scanning through we have depicted them in a
phase plot for polyelectrolytes using the parameters $\elj$ and
$\lb^{-1}$ as it was done for example in Ref.~\cite{schiessel98a}.
Some of the simulation series are marked with dashed lines in
fig.~\ref{phase_space1}. Close to the lines we have put some snapshots
of the configurations to get a visual impression.

Looking at series C3 crossing the entire $\lb$-range we find, starting
on the right side at the neutral case with $\lb=0.0\sigma$, a neutral
globule. The chain is collapsed due to the poor solvent. With
increasing $\lb$ the chain is getting charged, the self-energy grows,
and we can observe the Rayleigh instability at $\lb=0.75\sigma$ where
the globule splits into a dumbbell. At this point we find already some
counterions close to the chain, hence there are condensed counterions
present. Further increase of $\lb$ can lead to further Rayleigh
instabilities, depending on the parameters. Then, after reaching its
maximum extension at $\lb=1.25\sigma$ the chain slowly shrinks, since
the enlarged Coulomb repulsion gets overcompensated by condensing
counterions. Finally we reach a collapsed conformation at
$\lb=9.0\sigma$. The non-monotonic behavior of the chain extension is
qualitatively the same as in the good solvent case and qualitatively
well understood~\cite{schiessel98a,stevens95a,winkler97a}; however,
the decrease is faster and more pronounced in the poor solvent
case~\cite{micka99a,khan99a}.

Scaling theories have predicted that with the onset of condensation
the pearl-necklace should collapse in a first-order
transition~\cite{khokhlov80a,dobrynin96a,schiessel98a,vilgis00a}.
However we find always a smooth distribution of counterions which
looks like a distribution that can be calculated within PB
theory~\cite{deserno00a}. The counterions get pulled closer to the
macroion as the Coulomb coupling is increased, and a rather high
Bjerrum length, or similarly, a large number of condensed counterions
are needed to collapse the PE to a globule. Also the osmotic pressure
does not show any dramatic decrease with $\lb$, so we believe that for
an adequate description of the collapse we need a refined theory.

\begin{figure}[ht]  \begin{center}  
    \includegraphics[angle=0,width=0.95\linewidth]{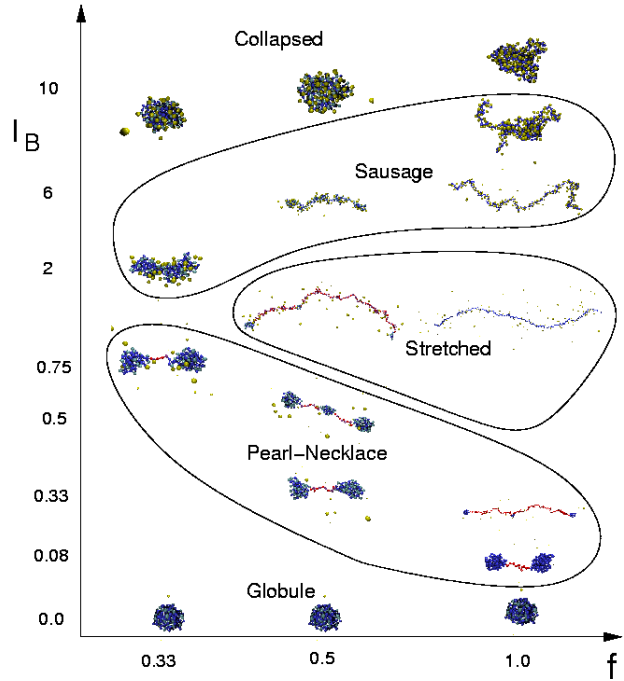}
    \caption{Schematic view of the location of the different PE
      configuration types in the $\lb$/$f$ phase diagram. The
      simulations are from series C1 at $f=1$, C2 at $f=\frac{1}{2}$
      and C3 at $f=\frac{1}{3}$ performed at $\elj=1.75\KB T$.
      \label{phase_space2}}
\end{center} \end{figure}

A very interesting, and as we believe, new conformational regime opens
up basically for those $\lb$ values between the maximal chain
extension up to the collapsed state. Here we find conformations that
are reminiscent of a cigar-like shape~\cite{khokhlov80a}, but turn out
to look more like a sausage for longer chains. Since with increasing
$\lb$ more counterions are attracted towards the chain the pearl pearl
repulsion is getting screened such that the pearls slowly coalesce and
the conformation is not stretched on longer length scales anymore. Due
to its shape we have termed it the sausage regime. Note that the
actual conformation depends also on $f$ where smaller values of $f$
lead to thicker sausages. Conformational snapshots are shown in
fig.\ref{phase_space2}.

At the crossing point of series B and C3 at $f=\frac{1}{3}$,
$\lb=1.5\sigma$ and $\elj=1.75\KB T$ being in the pearl-necklace
regime, we have studied the $\NM$-dependence of the chain
conformation. Starting from short chains which form a globule, e.g.
one pearl, we enter a number of Rayleigh instabilities upon
elongation. A few exemplary snapshots of series A1 have already been
shown in fig.~\ref{phase_spaceN}.

As a next step we have studied the dependence of the conformations
upon the charge fraction $f$. As one can see in
fig.~\ref{phase_space2} this opens up a completely new plane in the
phase diagram of polyelectrolytes. The three simulation series C1, C2
and C3 all scan the entire $\lb$-range and differ only in the charge
fraction $f$.

In mean field theories for polyelectrolytes we find two important
parameters. The first one is the Manning parameter $\xi = \lb b^{-1}
f$ which plays a role in all attempts to renormalize the charge of
highly charged polyelectrolytes~\cite{manning69a}. The second one is a
measure for the overall Coulombic repulsion on a charged chain $\lb
b^{-1} f^2$. As a first Ansatz for a scaling theory for highly charged
polyelectrolytes at finite density an effective charge fraction $\feff
= \frac{f}{\xi}$ is often used, trying to combine both parameters.  In
this framework the three simulation series should behave identically.
In contrast to this we find e.g. that the maximal extension $\RE
\mys{(max)}$ for the three series differs strongly.  Namely we find
for C1 $\RE \mys{(max)}=130\sigma$, for C2 $\RE
\mys{(max)}=66.4\sigma$ and for C3 $\RE \mys{(max)}=13.5\sigma$.  It
is also striking that only the series with larger $f$ has a regime
where the chains are stretched and they behave as if they were in a
good solvent.

The Rayleigh instability occurs at the same value $\lb b^{-1} f^2
\approx 1/12$ for all three series as expected because the counterions
do not play a dominant role here. Increasing $\lb$ from this point on
the three series behave very differently. Whereas the series with
$f=\frac{1}{2}$ and $f=1$ show a cascade of Rayleigh instabilities
budding more pearls until they reach stretched conformations the
series with $f=\frac{1}{3}$ has a dumbbell conformation at the maximum
extension.  The maximum extension itself is reached at different
values for $\lb$ (see fig.~\ref{fig.re_lbdep}).  For series C1 the
maximum extension is probably more restricted by the chain entropy
than that is determined by the interplay between repulsive and
attractive interactions. The chains then slowly shrink and enter the
sausage regime. Still at the same value of $\lb$ the chains with $f=1$
are much longer than the chains with $f=\frac{1}{2}$ or
$f=\frac{1}{3}$.  Finally the collapsed conformation is reached
roughly at the same value of $\lb \approx 8 \sigma$. In this regime we
have almost a dense electrolyte solution with mobile ions inside the
chains so we should be close to the critical behavior of a Coulomb
fluid. This suggest that the collapse should occur at roughly the same
value of the coupling parameter $v_m v_c \lb / \sigma$ which is the
interaction energy of two oppositely charged ions at contact in units
of $\KBT$.

In principle the phase space for polyelectrolytes has far more than
the shown three dimensions. As we know from previous
studies~\cite{micka99a,limbach01c,limbach02a,limbach02c,holm03a} also
the density is a very relevant parameter and would have to be included
into the phase diagram. Further important parameters are the valency
of the counterions, and added salt concentrations which we did not
investigate at all here. This all reflects the fact that there is
presumably no general parameter for the Coulomb interaction.

\section{Chain conformation and experiments \label{sec_formfac}}

The last section focuses on the connection between the chain
conformation and experimentally accessible observables like the
characteristic ratios $r$ and $\alpha$ and the form factor $S_1$.

\subsection{Characteristic ratios}

The characteristic ratios $r= (\RE/\RG)^2$ and $\alpha=\RG/\RH$ are
often used as a first step to characterize conformations that are
extended in one dimension. Since both $\RG$ and $\RH$ are accessible
with experimental methods it is interesting to know how $\alpha$ can
be used to distinguish different conformation types.

\begin{figure}[ht]  \begin{center}  
    \includegraphics[angle=0,width=0.95\linewidth]{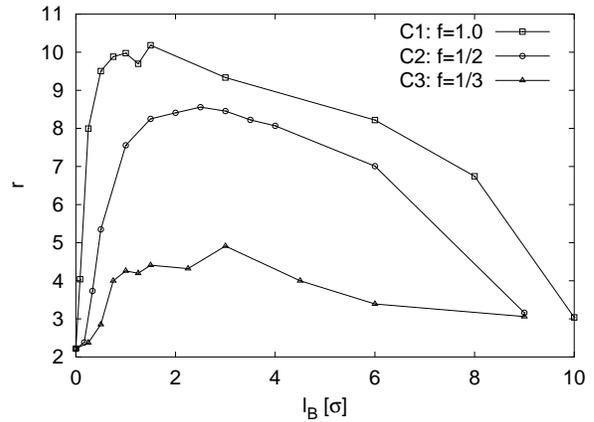}
    \caption{Change of the characteristic ratio $r$ with $\lb$ for
      different $f$ (series C1, C2 and C3).
      \label{fig_charr}}
\end{center} \end{figure}

\begin{figure}[ht]  \begin{center}  
    \includegraphics[angle=0,width=0.95\linewidth]{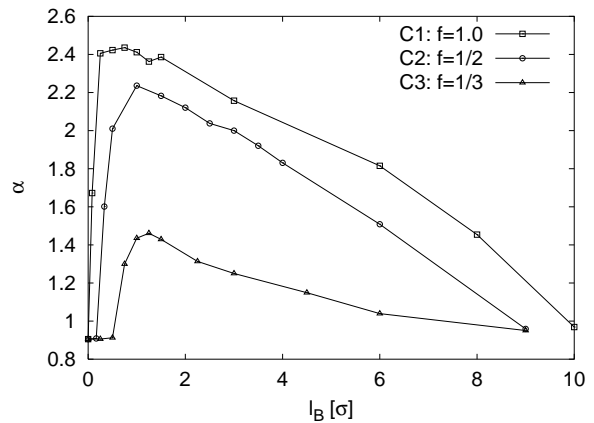}
    \caption{Change of the characteristic ratio $\alpha$ with $\lb$ for
      different $f$ (series C1, C2 and C3).
      \label{fig_alpha}}
\end{center} \end{figure}

In fig.~\ref{fig_charr} and fig.~\ref{fig_alpha} we show the change of
the characteristic ratios $r$ and $\alpha$ with $\lb$ for the series
C1, C2 and C3, e.g. for different $f$.
For the globular conformation at $\lb=0\sigma$ we find $r \approx 2.2$
and $\alpha \approx 0.9$. For dumbbells we find values for $r$ between
$4$ and $4.5$ and $\alpha \approx 1.6$. With further increase of $\lb$
both observables reach a maximum roughly at the maximal chain
extension (compare to fig~\ref{fig.re_lbdep}). Then $r$ and $\alpha$
decrease monotonically until the collapsed conformation is reached at
high values of $\lb$ where we find $r \approx 3$ and $\alpha \approx
1$. For a globular object one can calculate $r$ to be between $2$ and
$10/3$ which is consistent with our findings. For a completely
stretched object one would have $r=12$ which is of course not reached
with the simulations.
The form of the shown curves for $r$ and $\alpha$ clearly indicates
that it is not possible to deduce the conformation type since at each
value the chain could either be in a pearl-necklace conformation or,
at higher $\lb$ in a sausage like conformation.

\begin{figure}[ht]  \begin{center}  
    \includegraphics[angle=0,width=0.95\linewidth]{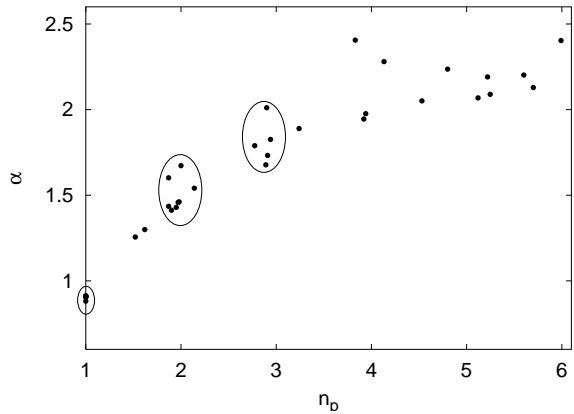}
    \caption{Characteristic ratio $\alpha$ for pearl-necklace
      structures as function of $\np$. The ellipses combine points
      close to $\np = 1,2 $ and $3$. 
      \label{fig_alpha_scat}}
\end{center} \end{figure}

Since for the pearl-necklace regime itself $\alpha$ is a monotonic
function of $\np$ we have plotted the values of $\alpha$ for all
simulations in this paper that we assign to this regime in
fig.~\ref{fig_alpha_scat}.
One can see a clear jump of $\alpha$ between $\np=1$ and $\np=2$. But
the range of $\alpha$ between conformations with two and three pearls
are already overlapping. With increasing $\np$ the slope of
$\alpha(\np)$ is decreasing and a further structure discrimination is
not possible.  So even in the pearl-necklace regime it is problematic
to use $\alpha$ to discriminate pearl-necklace conformations with
different numbers of pearls.

\subsection{Form factor}

\begin{figure}[ht]  \begin{center}   
    \includegraphics[angle=0,width=1.0\linewidth]{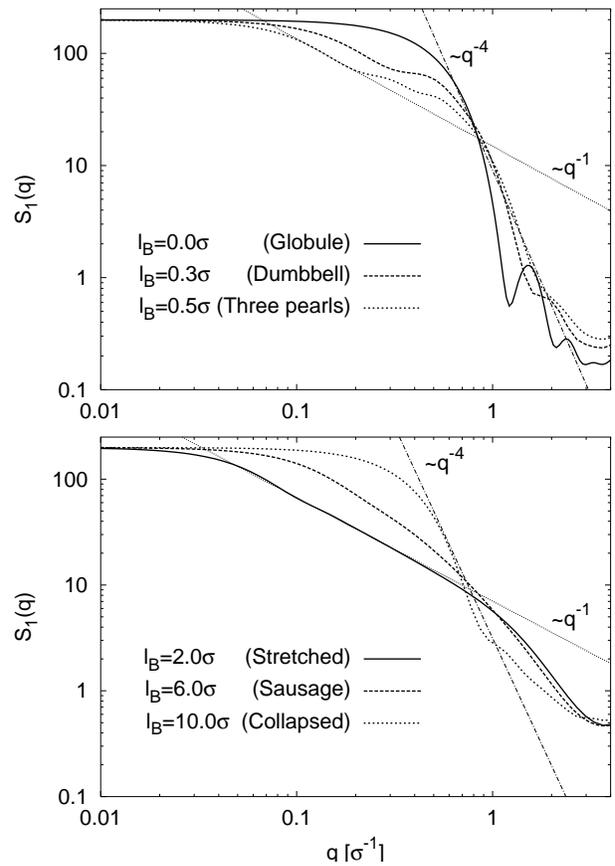}
    \caption{Form factors for the different conformation types while
      changing $\lb$ for series $C2$. The  straight lines indicate a
      stretched chain with $S_1 \propto q^{-1}$ and Porod scattering
      with $S_1 \propto q^{-4}$. \label{formfac_lbdep}}
\end{center} \end{figure}

More information about the chain conformation is contained in the form
factor $S_1$ (see eq.~\ref{formfac_sph}). In fig.~\ref{formfac_lbdep}
we show $S_1(q)$ for different simulations from series C2. The figure
shows the form factors for the different conformation types we have
found in our simulations, namely the neutral globule, the dumbbell,
the three pearl conformation, stretched chains, sausage like
conformations and a collapsed chain with most of the counterions being
inside.
The neutral globule shows Porod scattering. The strong oscillations
show that the globule has sharp boundaries and does not fluctuate
strongly. $S_1(q)$ for the dumbbell exhibits a shoulder at $q \approx
0.5\sigma^{-1}$ corresponding to a length of $\approx 12.5\sigma$ that
can be identified as the distance between the two pearls. The strong
decrease at higher $q$-values indicates again a Porod scattering in
this case coming from the surface of the pearls.  The minima of the
Porod scattering are smeared out by fluctuations of the shape and size
of the pearls. For the three pearl conformations we find two shoulders
one at $q \approx 0.2\sigma^{-1}$ and one at $q \approx
0.4\sigma^{-1}$. They are also indications of the pearl-pearl
distances. But the shoulders here are already less pronounced than in
the case of the dumbbell. The stretched conformation has a qualitative
different $S_1(q)$ showing a scaling with $q^{-1}$ over a large
$q$-range. It follows that the chain is stretched on length scales
larger than $10\sigma$ up to its full length of $66\sigma$. For the
sausage like conformation we can not identify any scaling regime or
other signatures. Thus we can not obtain another length scale than the
overall chain extension. The collapsed globule shows again, like the
neutral globule, a strong decrease of the scattering intensity at
large $q$-values. The rudimentary observable Porod scattering is
strongly smeared out showing that the competition between attractive
and repulsive forces induces large fluctuations on the surface of the
globular object.

After this overview over the scattering of the different conformation
types we discuss the form factor of pearl-necklace conformations with
a larger number of pearls in more detail.  In this case the scattering
is not only influenced by the fluctuations in shape and size of the
pearls and strings but also by the fluctuations of the structure type,
namely different numbers of pearls.

\begin{figure}[ht]  \begin{center}   
    \includegraphics[angle=0,width=0.95\linewidth]{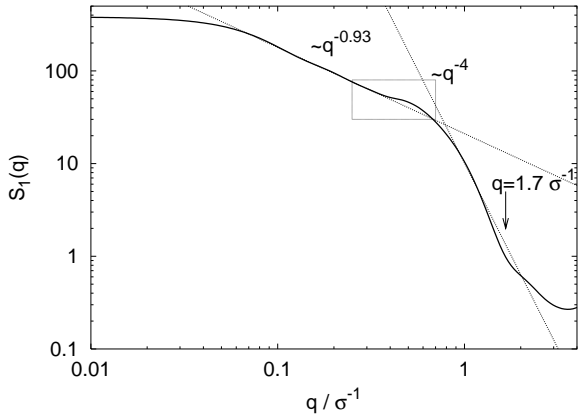}
    \caption{Form factor $S_1$ for typical
      pearl-necklace conformations. The dotted lines are fits to
      certain $q$-ranges, see text. The marked region is enlarged in
      fig.~\ref{formfac2}. The chains have a length $\NM=382$ and on
      average $\np=4.5$ pearls.  \label{formfac1}}
\end{center} \end{figure}

\begin{figure}[ht]  \begin{center}   
    \includegraphics[angle=0,width=0.95\linewidth]{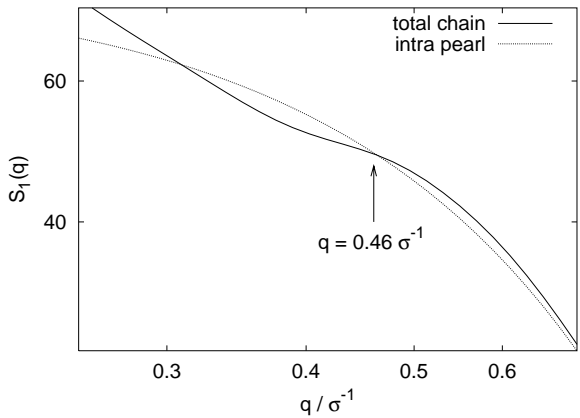}
    \caption{The pearl-pearl distance leaves only a very small
      signature in the form factor as can be seen in the close view of
      a comparison of the total form factor and the intra pearl
      form factor.\label{formfac2}}
\end{center} \end{figure}

The form factor shows four different regimes which reassemble the
different length scales that are involved in a pearl-necklace
structure. For chains taken from series A1 with $\NM=382$ the form
factor is plotted in fig.~\ref{formfac1}.
In the Guinnier regime at $\RG q \ll 1$ the radius of gyration $\RG$
can be calculated from $S_1(q)=\NM \left(1 - (\RG q)^{2}/3\right)$.
This yields $\RG=16.8\sigma \pm0.3\sigma$ in good agreement with the
directly calculated value $\RG=16.9\sigma\pm0.4\sigma$. In the
following range $0.07\sigma^{-1} \le q \le 0.4\sigma^{-1}$ the single
chain structure factor scales as $S_1(q)\propto q^{-1}$. The chain
conformations are thus stretched on length scales larger than
$15\sigma$. At $q \approx 0.5 \sigma^{-1}$ one can see a weakly
pronounced shoulder in $S_1$. A closer look to this region reveals
that $S_1(q)$ has an inflection point at $q=0.46 \sigma^{-1}$ which is
shown in fig.~\ref{formfac2}.
A comparison with the intra pearl scattering shows that this is due to
inter pearl scattering. This becomes more clear when one looks at the
analytic scattering function $S_a$ of a linear arrangement of $n$
homogenous spheres with a diameter $\rp$ and a distance $\rpp$.  $S_a$
is the product of the inter pearl scattering $S_{\mys{inter}}$ and the
intra pearl scattering $S_{\mys{intra}}$. The inter pearl scattering
is given by

\begin{equation}
S_{\mys{inter}} =  n + 2 \sum_{k=1}^{n-1} (n-k) \frac{\sin(q
    \rpp k)}{q \rpp k}  \mbox{ .}
\end{equation}

The intra pearl scattering is that of a homogenous sphere and thus
given by the Porod scattering:

\begin{equation}
S_{\mys{intra}} =  \frac{\sin(q \rp) - q \rp \cos(q \rp)}{q
    \rp^3} \label{eq.sa_intra}
\end{equation}

Dividing out the intra pearl scattering from $S_1$ gives access to the
inter pearl scattering and thus $\rpp$.  From the inflection point in
the inter pearl scattering one can calculate the pearl pearl distance.
This yields $\rpp=13.6\sigma$, again in accord with the directly
measured value $\rpp=13.3\sigma$. In the high $q$-range between $q
\approx 0.9 \sigma^{-1}$ and $q \approx 2.5 \sigma^{-1}$ we find
$S_1(q) \propto q^{-4}$, the typical Porod scattering with a small dip
at $q \approx 1.7 \sigma^{-1}$. Fitting the data to
eq.~\ref{eq.sa_intra} yields a pearl radius $\rp \approx 2.6 \sigma$
which again compares well to the directly calculated value $\rp
\approx 3.0 \sigma$.

We conclude that the cooperative effect of fluctuations on overlapping
length scales broadens all characteristic signatures which can be
revealed by scattering under experimental conditions like
polydispersity and line charge density fluctuations. Thus necklaces
might be difficult to detect. We find the most pronounced necklace
signatures for the dumbbell conformation.

\section{Conclusion}\label{sec:conclusion}

We have studied a dilute solution of strongly charged polyelectrolytes
in a poor solvent by means of molecular dynamics simulations for a
variety of different parameters.

We have developed a cluster algorithm to characterize pearl-necklace
structures even for single configurations and performed an extensive
data analysis of all the pearl-necklace conformations in our
simulations.

We found that the range where scaling predictions are applicable is
confined to a small range of parameters. One either has to stay close
to the infinite dilution limit or use weakly charged chains to exclude
counterion effects. The reason is that even in a dilute solution there
is already a delicate interplay between the counterion distribution
and the chain conformation for strongly charged chains.  This became
in particular evident in our investigations studying the dependence on
the solvent quality and the Coulomb parameters $\lb$ and $f$.

Our results challenge approaches that apply Manning-like charge
renormalization and simple scaling concepts to strongly charged
polyelectrolytes. Since we found strong effects already for dilute
solutions we expect an even more complicated counterion chain coupling
for semi-dilute and dense polyelectrolyte solutions.

We have discussed the different types of fluctuations present for
pearl-necklace structures and quantified them for our data. We showed
that the free energy differences between the structure types can be
very small and that the size and position of the substructures exhibit
large fluctuations.

In a preliminary classification of the phase space of PEs in poor
solvent we have shown that the phase space is multi-dimensional and
that the region for pearl-necklace structures is rather small. In this
context we have also investigated the Coulomb induced collapse of poor
solvent PEs which appears to be a rather smooth transition. There we
have shown that the strong screening in the vicinity of the charged
chains leads to sausage like conformations instead of pearl-necklaces.
Our data suggest that the collapse is determined by the Bjerrum length
since it occurs at $\lb \approx 8\sigma$ for all systems which lead us
to suggest that in the collapsed state ion-ion correlations $\propto
\lb$ play the most prominent part.

Finally we computed some experimentally accessible observables like
the form factor and characteristic size ratios. This should help to
analyze the experimental data and to discover pearl-necklace
signatures which, according to our understanding, are not very
pronounced.

\section{Acknowledgments}
We thank K. Kremer, B. D{\"u}nweg, B. Mergell, H. Schiessel and M. N.
Tamashiro for many fruitful discussions and comments.  We also want to
thank R. Everaers for contributions to the cluster algorithm. We
gratefully acknowledge partial funding by the DFG SPP 1009 and SFB
625, and through the ``Zentrum f\"ur Multifunktionelle Werkstoffe und
Miniaturisierte Funktionseinheiten", grant BMBF 03N 6500.


\appendix 
\begin{appendix}

\section{Simulated Systems}

Here we give a detailed overview over the parameters of the simulated
systems and some basic observables, namely the chain length $\NM$, the
end-to-end distance $\RE$, the hydrodynamic radius $\RH$, the
characteristic ratios $r= (\RE / \RG)^2$ and $\alpha= \RG / \RH$ and
the osmotic coefficient $OC=\Pi / p_{ig}$, where $\Pi$ is the pressure
and $p_{ig}$ is the ideal gas pressure. If possible, also the number
of pearls $\np$ and the average number of monomers in a pearl $\gp$ is
given. In the tables~\ref{serieA1_tab}~to~\ref{serieD_tab} $\RE$,
$\RH$ and $\lb$ are given in $\sigma$, $\elj$ in $\KB T$.  All other
quantities in the tables are dimensionless numbers.  The statistical
error for the chain extensions $\RE$ and $\RH$ is smaller than $10\%$.
$\gp$ exhibits a systematic error of $\pm4$ monomers. For $\gp > 30$
the number of pearls has an accuracy of $\approx 5\%$. For smaller
$\gp$ values this error becomes larger. The statistical error of the
pressure calculation gives rise to an error of the osmotic coefficient
of $\pm0.06$ resulting in a large relative error for small values of
$OC$.

\begin{table}[ht] 
  \caption{ {\bf Series A1: $\NM = 48 ... 478$, } \hspace{0.2\linewidth}
    $\elj=1.75\KB T$, $\lb=1.5\sigma$, $f=1/3$, $\rhoc=1.0\times
    10^{-5}\sigma^{-3}$ \label{serieA1_tab}} 
  \begin{center} \small
    \begin{tabular}{c|cc|cc|cc|c} 
      $\NM$& $\RE$& $\RH$& $r$  &$\alpha$&$np$& $\gp$& OC \\ \hline
      48   & 3.48 & 2.19 & 3.25 & 0.88 & 1.00 & 46.0 & 0.96\\ 
      94   & 4.56 & 2.71 & 3.40 & 0.91 & 1.01 & 93.9 & 0.68\\ 
      142  & 12.9 & 4.11 & 4.63 & 1.46 & 1.97 & 68.8 & 0.60\\ 
      190  & 15.0 & 4.60 & 4.49 & 1.54 & 2.01 & 90.4 & 0.44\\ 
      238  & 25.7 & 5.86 & 5.84 & 1.83 & 2.94 & 75.5 & 0.46\\ 
      286  & 30.4 & 6.51 & 6.10 & 1.89 & 3.24 & 82.5 & 0.42\\ 
      334  & 38.2 & 7.44 & 6.73 & 1.98 & 3.94 & 78.6 & 0.40\\ 
      382  & 45.4 & 8.24 & 7.24 & 2.05 & 4.53 & 78.0 & 0.35\\ 
      430  & 55.4 & 9.22 & 7.55 & 2.19 & 5.22 & 75.2 & 0.36\\ 
      478  & 59.7 & 9.81 & 7.61 & 2.20 & 5.60 & 78.3 & 0.39\\ 
    \end{tabular}
  \end{center}
\end{table}

\begin{table}[ht] 
     \caption{ {\bf Series A2: $\NM = 100 ... 300$, }  \hspace{0.2\linewidth}
$\elj=1.75\KB T$, $\lb=1.5\sigma$, $f=1/2$, $\rhoc=6.7\times
10^{-5}\sigma^{-3}$ \label{serieA2_tab}} 
  \begin{center} \small
    \begin{tabular}{c|cc|cc|cc|c} 
      $\NM$&$\RE$& $\RH$& $r$  &$\alpha$&$np$& $\gp$& OC \\ \hline
      99   &26.8 &5.59  &7.17 &1.79 &2.77 &23.3 &0.54\\ 
      199  &68.1 &10.43 &8.20 &2.28 &4.13 &20.7 &0.32\\ 
      299  &102.9&14.37 &8.88 &2.40 &5.99 &18.9 &0.39\\ 
    \end{tabular}
  \end{center}
\end{table}

\begin{table}[ht] 
\caption{
{\bf Series A3: $\NM = 100 ... 300$, } \hspace{0.2\linewidth}
$\elj=1.75\KB T$, $\lb=1.5\sigma$, $f=1/2$, $\rhoc=6.7\times
10^{-5}\sigma^{-3}$ \label{serieA3_tab}
} 
  \begin{center} \small
    \begin{tabular}{c|cc|cc|cc|c} 
      $\NM$&$\RE$& $\RH$& $r$  &$\alpha$&$np$& $\gp$& OC \\ \hline
      99   &22.1 & 5.08 & 6.72 & 1.68   &2.89& 25.1 & 0.46 \\ 
      199  &54.9 & 9.15 & 8.24 & 2.09   &5.25& 21.4 & 0.36 \\ 
      299  &83.4 & 12.6 & 8.69 & 2.24   &7.68& 20.4 & 0.35 \\ 
    \end{tabular}
  \end{center}
\end{table}

\begin{table}[ht]  
  \caption{{\bf Series B: $\elj=0.0\KB T ... 2.0\KB T$, }\hspace{0.2\linewidth}
    $\NM=238$, $\lb=1.5\sigma$, $f=1/3$ , $\rhoc=5.0\times 10^{-5}\sigma^{-3}$\label{serieB_tab}}
 \begin{center} \small  
    \begin{tabular}{c|cc|cc|cc|c} 
      $\elj$&$\RE$& $\RH$& $r$  &$\alpha$& $np$ & $\gp$& OC \\ \hline
      0.00  &102  & 16.2 & 8.88 &  2.11  & ---  & ---  & 0.66 \\ 
      0.50  &100  & 14.9 & 9.28 &  2.21  & ---  & ---  & 0.62 \\ 
      1.00  &82.2 & 13.1 & 8.81 &  2.11  & ---  & ---  & 0.63 \\ 
      1.25  &68.7 & 11.0 & 8.44 &  2.15  & ---  & ---  & 0.53 \\ 
      1.35  &58.5 & 9.78 & 7.91 &  2.13  & 5.70 & 25.2 & 0.52 \\ 
      1.45  &45.9 & 8.22 & 7.32 &  2.07  & 5.12 & 35.6 & 0.53 \\ 
      1.55  &34.3 & 6.89 & 6.56 &  1.94  & 3.92 & 52.9 & 0.45 \\ 
      1.65  &24.2 & 5.83 & 5.76 &  1.73  & 2.91 & 76.8 & 0.48 \\ 
      1.75  &16.8 & 5.03 & 4.72 &  1.54  & 2.14 & 108  & 0.36 \\ 
      1.85  &13.8 & 4.68 & 4.36 &  1.41  & 1.90 & 123  & 0.31 \\ 
      2.00  &11.2 & 4.34 & 4.21 &  1.26  & 1.52 & 155  & 0.26 \\ 
    \end{tabular}
  \end{center} 
\end{table}

\begin{table}[ht]  
  \caption{ {\bf Series C1: $\lb = 0.0 ... 12.0 \sigma$, } \hspace{0.2\linewidth}
$f=1$, $\elj=1.75\KB T$, $\NM=200$, $\rhoc=5.0\times
10^{-5}\sigma^{-3}$ \label{serieC1_tab}}
  \begin{center}    \small
    \begin{tabular}{c|cc|cc|cc|c} 
      $\lb$&$\RE$ &$\RH$& $r$  &$\alpha$&$np$& $\gp$& OC \\ \hline
      0.0  & 4.41 & 3.27& 2.22 &  0.91  &1.0 & 200  & 0.97 \\ 
      0.083& 15.8 & 4.70& 4.04 & 1.67   &2.0 & 95.3 & 0.94 \\ 
      0.25 & 72.1 & 10.6& 7.99 & 2.41   &3.83& 21.1 & 0.86 \\ 
      0.5  & 111.3& 14.9& 9.51 & 2.42   & ---& ---  & 0.75 \\ 
      0.75 & 124.8& 16.3& 9.88 & 2.44   & ---& ---  & 0.63 \\ 
      1.0  & 129.5& 17.0& 9.98 & 2.41   & ---& ---  & 0.55 \\ 
      1.25 & 125.8& 17.1& 9.70 & 2.36   & ---& ---  & 0.46 \\ 
      1.5  & 130.2& 17.1& 10.2 & 2.39   & ---& ---  & 0.40 \\ 
      3.0  & 104.8& 15.9& 9.34 & 2.16   & ---& ---  & 0.18 \\ 
      6.0  & 67.66& 13.0& 8.22 & 1.82   & ---& ---  & 0.06 \\ 
      8.0  & 31.94& 8.46& 6.74 & 1.45   & ---& ---  & 0.06 \\ 
      10.0 & 9.204& 5.45& 3.04 & 0.97   & 1.0& 200  & 0.04 \\ 
    \end{tabular}
  \end{center} 
\end{table}

\begin{table}[ht]  
  \caption{{\bf Series C2: $\lb = 0.0 ... 6.0 \sigma$, }  \hspace{0.2\linewidth}
$f=1/2$, $\elj=1.75\KB T$, $\NM=199$, $\rhoc=5.0\times
10^{-5}\sigma^{-3}$ \label{serieC2_tab}}
  \begin{center}    \small
    \begin{tabular}{c|cc|cc|cc|c} 
      $\lb$&$\RE$ &$\RH$& $r$  &$\alpha$&$np$& $\gp$& OC \\ \hline
      0.0  & 4.41 & 3.27& 2.22 &  0.91  &1.0 & 199  & 0.97 \\ 
      0.167& 4.63 & 3.30& 2.39 &  0.91  &1.0 & 199  & 0.93 \\ 
      0.333& 14.2 & 4.59& 3.72 &  1.60  &1.87& 102  & 0.82 \\ 
      0.5  & 26.6 & 5.72& 5.38 &  2.00  &2.90& 61.4 & 0.79 \\ 
      1.0  & 54.7 & 8.90& 7.53 &  2.24  &4.80& 25.0 & 0.57 \\ 
      1.5  & 65.2 & 10.4& 8.22 &  2.19  & ---& ---  & 0.45 \\ 
      2.0  & 66.4 & 10.8& 8.40 &  2.12  & ---& ---  & 0.36 \\ 
      2.5  & 62.6 & 10.5& 8.52 &  2.04  & ---& ---  & 0.27 \\ 
      3.0  & 59.9 & 10.3& 8.48 &  2.00  & ---& ---  & 0.21 \\ 
      3.5  & 54.2 & 9.84& 8.26 &  1.92  & ---& ---  & 0.16 \\ 
      4.0  & 48.0 & 9.23& 8.08 &  1.83  & ---& ---  & 0.13 \\ 
      6.0  & 27.8 & 6.96& 7.03 &  1.50  & ---& ---  & 0.06 \\ 
      9.0  & 7.45 & 4.37& 3.16 &  0.96  &1.14& 173  & 0.05 \\ 
    \end{tabular}
   \end{center} 
\end{table}

\begin{table}[ht]  
  \caption{ {\bf Series C3: $\lb = 0.0 ... 9.0 \sigma$, } \hspace{0.2\linewidth}
$f=1/3$, $\elj=1.75\KB T$, $\NM=199$, $\rhoc=5.0\times
10^{-5}\sigma^{-3}$ \label{serieC3_tab}}
  \begin{center}  \small
    \begin{tabular}{c|cc|cc|cc|c} 
      $\lb$&$\RE$& $\RH$& $r$  &$\alpha$&$np$& $\gp$& OC \\ \hline
      0.0  &4.41 & 3.27 & 2.22 &  0.91  &1.0 & 199  & 0.97 \\ 
      0.25 &4.59 & 3.29 & 2.37 &  0.91  &1.0 & 199  & 0.93 \\ 
      0.5  &5.10 & 3.31 & 2.86 &  0.91  &1.0 & 199  & 0.76 \\ 
      0.75 &10.5 & 4.04 & 4.01 &  1.30  &1.62& 120  & 0.64 \\ 
      1.0  &13.0 & 4.39 & 4.24 &  1.44  &1.87& 104  & 0.54 \\ 
      1.25 &13.5 & 4.51 & 4.22 &  1.46  &1.98& 97.4 & 0.47 \\ 
      1.5  &13.5 & 4.50 & 4.38 &  1.43  &1.95& 99.2 & 0.38 \\ 
      2.25 &12.2 & 4.47 & 4.32 &  1.31  &1.57& 125  & 0.29 \\ 
      3.0  &12.3 & 4.44 & 4.90 &  1.25  &1.30& 152  & 0.22 \\ 
      4.5  &9.86 & 4.29 & 3.99 &  1.25  &1.06& 188  & 0.10 \\ 
      6.0  &7.77 & 4.06 & 3.39 &  1.04  &1.01& 198  & 0.15 \\ 
      9.0  &6.35 & 3.82 & 3.07 &  0.95  &1.00& 199  & 0.07 \\ 
    \end{tabular}
  \end{center} 
\end{table}

\begin{table}[ht]  
  \caption{ {\bf Series C4: $\lb = 1.5 ... 12.0 \sigma$, } \hspace{0.2\linewidth}
$f=1/3$, $\elj=1.5\KB T$, $\NM=94$, $\rhoc=1.0\times
10^{-5}\sigma^{-3}$ \label{serieC4_tab}}
  \begin{center}  \small
    \begin{tabular}{c|cc|cc|cc|c} 
      $\lb$&$\RE$& $\RH$& $r$  &$\alpha$&$np$& $\gp$& OC \\ \hline
      1.5 &13.5 &3.93 &5.33 &1.49 &2.04 &41.2 &0.68\\
      1.8 &13.9 &4.00 &5.47 &1.48 &2.10 &39.6 &0.62\\
      2.1 &15.1 &4.15 &5.88 &1.50 &2.24 &36.2 &0.59\\
      2.4 &15.7 &4.23 &6.14 &1.50 &2.27 &35.3 &0.49\\
      2.7 &15.6 &4.22 &6.41 &1.46 &2.26 &35.9 &0.53\\
      3.0 &15.6 &4.23 &6.47 &1.45 &2.26 &35.8 &0.40\\
      3.3 &14.4 &4.10 &6.43 &1.39 &2.17 &38.5 &0.45\\
      3.6 &13.2 &3.97 &6.22 &1.33 &2.06 &41.6 &0.30\\
      3.9 &12.2 &3.85 &6.17 &1.28 &1.92 &45.6 &0.33\\
      5.1 &8.6  &3.44 &5.08 &1.10 &1.38 &66.7 &0.22\\
      5.4 &8.2  &3.39 &5.06 &1.08 &1.30 &71.2 &0.17\\
      5.7 &8.0  &3.37 &4.96 &1.07 &1.26 &73.6 &0.08\\
      6.0 &7.7  &3.34 &4.81 &1.05 &1.22 &76.4 &0.12\\
      9.0 &5.7  &3.14 &3.58 &0.97 &1.03 &91.4 &0.19\\
      12.0 &5.1 &3.06 &3.17 &0.94 &1.01 &93.4 &0.04\\          
    \end{tabular}
  \end{center} 
\end{table}

\begin{table}[ht]  
  \caption{ {\bf Series D:  $\lb b^{-1} f^2 = 0.25=$const., } \hspace{0.2\linewidth}
$\elj=1.75\KB T$, $\NM=200$, $\rhoc=5.0\times 10^{-4}\sigma^{-3}$ \label{serieD_tab}}
  \begin{center} \small
    \setlength{\tabcolsep}{1ex}
    \begin{tabular}{cc|cc|cc|cc|c} 
      $\lb$&$f$&$\RE$& $\RH$& $r$  &$\alpha$&$np$& $\gp$& OC \\ \hline
      0.25 &1.00 &59.4 &9.54 &7.81 &2.23 &4.88 &21.6 &0.75\\
      1.00 &0.50 &32.3 &6.66 &6.96 &1.84 &4.37 &39.2 &0.44\\
      2.25 &0.33 &8.3  &4.15 &2.90 &1.18 &1.11 &179&0.17\\
      4.00 &0.25 &5.9  &3.55 &3.00 &0.96 &1.00 &201&0.13\\
    \end{tabular}
  \end{center} 
\end{table}

\end{appendix}

\end{document}